\documentclass[usenatbib,letterpaper]{mn2e}
\usepackage{epsfig}

\newcommand{\yr}{\,\mbox{yr}}
\newcommand{\Myr}{\,\mbox{Myr}}
\newcommand{\Gyr}{\,\mbox{Gyr}}

\newcommand{\kpc}{\,\mbox{kpc}}
\newcommand{\kms}{\,\mbox{km}\,\mbox{s}^{-1}}

\newcommand{\Msun}{\,M_{\sun}}
\newcommand{\msun}{\,M_{\sun}}
\newcommand{\degree}{^\circ}

\newcommand{\tsim}{\sim\!}
\newcommand{\ea}{et al.~}
\newcommand{\gtrsim}{\ga}
\newcommand{\ltrsim}{\la}
\newcommand{\araa}{ARA\&A}
\newcommand{\apj}{ApJ}
\newcommand{\apjl}{ApJ}
\newcommand{\mnras}{MNRAS}
\newcommand{\aj}{AJ}

\newcommand{\aap}{A\&A}
\defcitealias{geehan06}{Paper~I}
\defcitealias{fardal06}{Paper~II}

\title[Investigating the Andromeda Stream: III] 
  {Investigating the Andromeda Stream: III.  A Young Shell System in M31}
\author[M. A. Fardal et al.]       
{M. A. Fardal$^1$\thanks{E-mail: fardal@fcrao1.astro.umass.edu,  
  raja@ucolick.org, babul@uvic.ca, alan@uvic.ca},                
  P. Guhathakurta$^2$, A. Babul$^3$, and A. W. McConnachie$^3$ \\
  $^1$Dept.\ of Astronomy, University of Massachusetts, 
      Amherst, MA, 01003, USA \\
  $^2$UCO/Lick Observatory, Dept.\ of Astronomy \& Astrophysics,
      Univ. of California, 1156 High St., Santa Cruz, CA, 95064, USA \\
  $^3$Dept.\ of Physics \& Astronomy, University of Victoria, 
      Elliott Building, 3800 Finnerty Rd., Victoria, BC, V8P 1A1, Canada}

\date{draft version \today}
\pagerange{\pageref{firstpage}--\pageref{lastpage}} \pubyear{2006} 
\begin{document}
\maketitle  
\label{firstpage}
\begin{abstract}
Published maps of red giant stars in the halo region of M31 exhibit a
giant stellar stream to the south of this galaxy, as well as a giant
``shelf'' to the northeast of M31's center.  Using these maps, we find
that there is a fainter shelf of comparable size on the western side
as well.  By choosing appropriate structural and orbital parameters for 
an accreting dwarf satellite within the accurate M31 potential model of
Geehan \ea (2006), we produce a very similar structure in an $N$-body
simulation.  In this scenario, the tidal stream produced at pericenter
of the satellite's orbit matches the observed southern stream, while
the forward continuation of this tidal stream makes up two orbital
loops, broadened into fan-like structures by successive pericentric
passages; these loops correspond to the NE and W shelves.  The tidal
debris from the satellite also reproduces a previously-observed
``stream'' of counterrotating PNe and a related stream seen in red
giant stars.  The debris pattern in our simulation resembles the shell
systems detected around many elliptical galaxies, though this is 
the first identification of a shell system in a spiral galaxy 
and the first in any galaxy close enough to allow measurements of stellar
velocities and relative distances.  We discuss the physics of these
partial shells, highlighting the role played by spatial and velocity
caustics in the observations.  We show that kinematic surveys of the
tidal debris will provide a sensitive measurement of M31's halo
potential, while quantifying the surface density of debris in the
shelves will let us reconstruct the original mass and time of
disruption of the progenitor satellite.
\end{abstract}

\section{INTRODUCTION}
Our neighboring disk galaxy M31 has turned out to be a well-equipped
laboratory for research into the accretion and disruption of
substructure in galactic halos.  A stellar stream of length $\tsim 120 \kpc$ 
is observed to flow from its tip south and behind M31 into M31's center
\citep{ibata01,mcconnachie03,ibata04,raja06}, while many other irregular
morphological and kinematic features indicative of accretion are seen
in the galactic disk and halo \citep{ferguson02,merrett03,ibata05}.
The giant southern stream is thought to be formed at the last pericentric
passage of a progenitor satellite with mass $\tsim 10^9 \Msun$ on a
highly radial orbit (\citealp{ibata04,font06}; \citealp{geehan06}, 
hereafter Paper~I; \citealp{fardal06}, hereafter Paper~II).  
This stream holds out the promise of the first precise 
measurements of the gravitational potential in M31's halo.  
This is important for many reasons, including breaking the disk-halo 
degeneracy in the mass distribution \citepalias{geehan06}
and thereby understanding the response of the dark halo to the 
central baryons.  However, there are two obstacles to these measurements 
at present \citepalias{fardal06}.  First, the current observational errors on the
distance allow too much flexibility in the orbits to constrain the
potential effectively.  Second, the stream does not follow a single
orbit, but has a gradient of energy along the length of the stream,
``tilting'' the stream in phase space relative to the orbit and
biasing the measurement of the potential.  The amount of this tilt
depends on the location of the progenitor, which is currently unknown.
Locating the progenitor would both specify the stream-orbit tilt and
constrain the allowed orbital trajectories, removing both obstacles.

The ``Northeast Shelf'', a faint diffuse feature seen to the NE side of
M31 in the red giant branch (RGB) star-count maps of
\citet{ferguson02}, has been suggested as a possibility for the
current location of the progenitor, or at least as the forward
continuation of the stream \citep{ibata04,font06}.  Below we find 
that there is a similar, even fainter shelf or plateau on the other
side of M31, which we call the ``Western shelf''.  In this paper, we
focus on the possibility that the forward continuation of the stream
makes up these NE and W shelves.  In this scenario, the shelves are a
similar phenomenon to the ``shell'' systems seen around many
elliptical galaxies.  These systems are believed to be produced by
tidal disruption of a satellite galaxy on a nearly radial orbit
\citep[e.g.,][]{schweizer80,quinn84,hernquist88,barnes92}, exactly the
phenomenon believed to be responsible for the creation of M31's
southern stream, so it is not unexpected that shell-like features
should be present. These coherent structures are rich in information.
For example, they allow sensitive tests of the potential of the
host galaxy without the need for distance information \citep{merrifield98}, 
if the necessary kinematic observations can be obtained as is possible for 
the first time in the M31 shells.  Also, there exists a simple relation 
between the surface density of a shell and the rate at which stellar mass 
flows through the shell; when appied to the W shelf, this can be used to 
deduce the orbital parameters and the mass of the satellite that produced 
the observed debris.

Our purpose in this paper is threefold: first, to show that this 
paradigm of a shell debris pattern
is consistent with the current observations of M31; second, to
explain the basic physics that gives rise to the spatial and kinematic
pattern of the debris; and third, to find observational diagnostics 
to further constrain the model of the debris from present and future
datasets.  
In \S~2, we give an overview of the complex observed structure around
M31, and describe our scenario to explain a major part of this structure.  
In \S~3, we explain our methods for setting up and running a
simulation of this scenario.  We then compare the simulation results
to the observed properties of the debris around M31, to see whether
our model is plausible.  In \S~4, we discuss the physics that governs
this model of the satellite debris.  We consider several observables
that can be used to confirm or rule out our scenario, and to constrain
the physical properties of the stream and its progenitor.  In \S~5, we
discuss some remaining issues regarding our model, and state our
conclusions.

\nobreak

\section{SCENARIO OF SHELF FORMATION}

\begin{figure*}
\begin{center}
\leavevmode \epsfysize=9cm \epsfbox{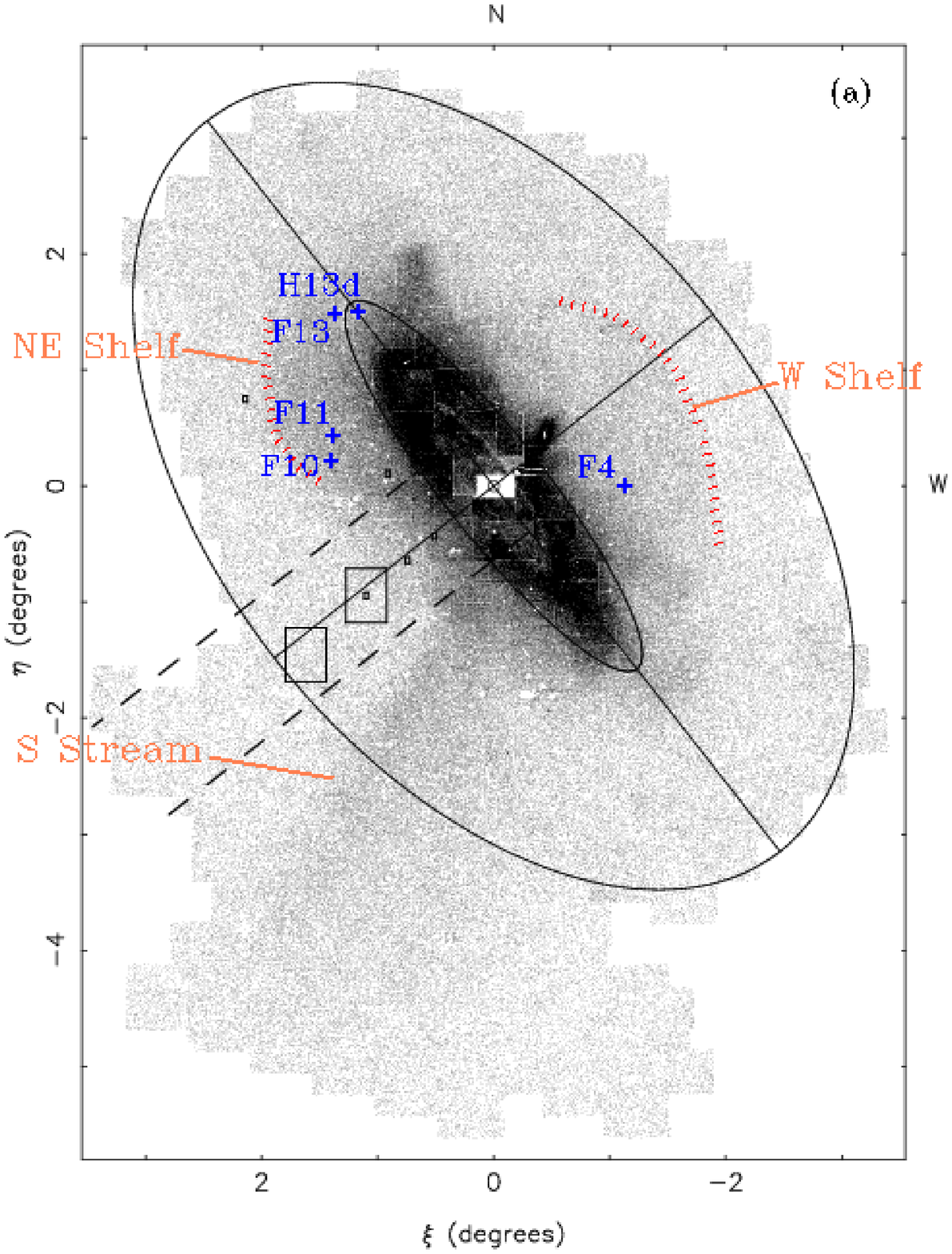} \epsfysize=9cm \epsfbox{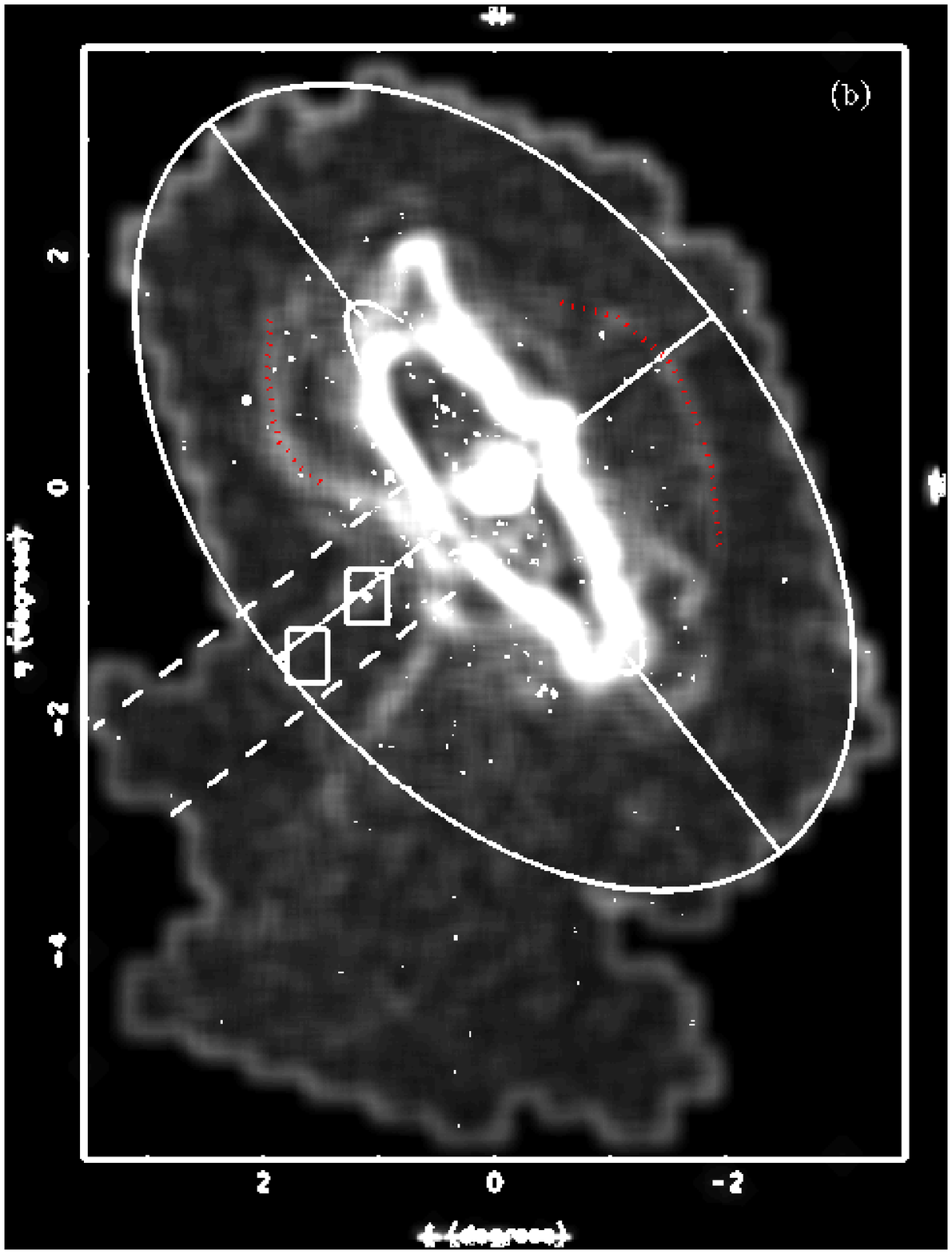}
\caption{
\label{fig.sky}
{\it Panel a:} Map of RGB count density, from \citet{irwin05}.  
The edges of the NE and W shelves that are the focus of this
paper are marked with red dotted lines. 
The H13d field of \citet{reitzel06} and 
four fields from \citet{ibata05} discussed
below are marked with crosses.
{\it Panel b:} Sobel-filtered version of Panel a, 
which detects sharp edges in the count map.  To create this map,
we fill in sharp features such as plot annotations and noise
spikes with the surrounding pixel values, smooth the map, 
and apply the Sobel operator.   We then fit a smooth curve to
the shelf edges, which again are shown by dotted lines.
}
\end{center}
\end{figure*}

Figure~\ref{fig.sky}a shows the density of RGB stars around M31 (from
\citealp{irwin05}).  The sky coordinates $\xi$ and $\eta$ in the
eastern and northern (E and N) directions respectively are ``standard
coordinates'' centered on M31.  Several features that will be
important in the discussion below are marked, including the southern
(S) stream.  The overdensity known as the NE shelf
\citep{ferguson02,ferguson05} is also readily apparent.

Less obviously, there is a similar enhancement of lower surface
brightness on the W side, which is apparent on close visual inspection
in this map and several of its earlier incarnations
\citep{ferguson02,lewis04,mcconnachie04news,ferguson05}.  To enhance this
structure, we display in Figure~\ref{fig.sky}b a map obtained from 
Figure~1a using
the Sobel edge-detection operator, which turns sharp edges into bright
ridges.  We fit the NE and W shelf ridges in this map with smooth
curves, whose interpolating points are given in
Table~\ref{table.edges}.  The fainter W shelf has not previously been
been singled out as an interesting feature in the literature, 
but it appears clearly enough here to investigate its possible origins.

Several other interesting features show up in Figure~\ref{fig.sky}b, 
among them the ``Northern Spur'', a bright loop at $\xi
\approx 0.8 \degree$, $\eta \approx 1.8 \degree$; the ``G1 clump'' and
related structure near the SW major axis at the outer regions of the
disk; and a shelf in the southern stream at $\eta \approx -1.0
\degree$.  Whether these other features have any relation to the
stream itself is not currently known.

\begin{table}
\caption{Position of shelf edges (dotted lines in
Figure \protect\ref{fig.sky}).}
\label{table.edges}
\begin{tabular}{ccc}
\hline
$\xi$~($\degree$) & $\eta$~($\degree$) & $R_{proj}$~($\degree$) \\
\hline
NE shelf & \\
\hline
1.95 &  1.46   &  2.43	\\
1.96 &  0.93   &  2.17	\\
1.91 &  0.59   &  2.00	\\
1.86 &  0.41   &  1.91	\\
1.78 &  0.31   &  1.81	\\
1.70 &  0.17   &  1.71	\\
1.45 & $-0.01$ &  1.45	\\
\hline
W shelf & \\
\hline
$-0.56$ &   1.61  &  1.70  \\
$-1.03$ &   1.49  &  1.81  \\
$-1.45$ &   1.16  &  1.85  \\
$-1.58$ &   0.97  &  1.86  \\
$-1.78$ &   0.47  &  1.84  \\
$-1.87$ &   0.17  &  1.87  \\
$-1.92$ & $-0.31$ &  1.94  \\
$-1.95$ & $-0.56$ &  2.03  \\
\hline
\end{tabular}
\end{table}

In \citetalias{fardal06}, we found that the southern stream is well
explained by a baryon-dominated dwarf satellite of mass $\tsim 10^9
\Msun$, experiencing its first pass through M31 pericenter.
Generically, in such a collision, much of the satellite's mass is
stripped and turns into leading and trailing tidal streams.  These two
streams are roughly symmetric about the progenitor, though the leading
stream spirals in towards the primary galaxy and the trailing stream
outwards from it, as one traces the streams away from the progenitor.
In M31, the trailing, higher-energy stream is the one that must
correspond to the observed southern stream, since the surface density
is observed to fall off towards the tip of this stream opposite to its
direction of motion.  The locations of the progenitor itself and of
the forward, lower-energy stream are currently unknown.  If the
satellite should still be intact at its next pericentric passage, it
again produces two streams, which nearly overlay the portions of the
original stream nearest the satellite.  A clear example of such a 
multiple stream system is shown in \citet{law05} using a simulation of 
the Sagittarius galaxy.

Since the progenitor is not observed along M31's southern stream,
which spans a range of $\tsim 0.4$ in orbital phase (measured in units
of the radial period), it is likely that the stream continues at least
an equal amount of 0.4 periods ahead, taking it past the next pericenter
(the second in our numbering scheme) and {\em at least}
a significant part of the way to the next apocenter.  In
\citetalias{fardal06}, we could not trace the continuation of the 
southern stream precisely past the second pericenter where it merges
into M31's bulge and disk.  However, we found that it was
constrained to lie on the NE side of M31.  We found an acceptable (if
not optimal) orbital fit putting the progenitor in the area of the NE
shelf.  In contrast, the progenitor would need one more
pericentric passage (its third) to reach the W shelf, as would 
any material in the tidal stream.  We also see in Figure~1
that the surface density of the W shelf is lower than for the NE shelf,
as if the density were tailing off in the stream forward of the
progenitor.

These facts lead us to our basic scenario: as we follow the
continuation of the southern stream further forward in the orbital
trajectory, the debris first produces a diffuse shelf on the E side of
M31, and subsequently a similar shelf on the W side.  Depending on the
current position and state of the progenitor, the NE shelf may be made
up solely of material from the forward continuation of the S stream,
or include the remnant of the progenitor and the streams created at
its second pericentric passage as well.  In the next section, we test
this scenario.

\section{SIMULATION OF DEBRIS PATTERN VERSUS OBSERVATIONS}
\subsection{Method}
In Papers I and II, we discussed the 
physical geometry of M31 and its
surrounding system of satellites and streams.  We then derived a
bulge-disk-halo model for M31's gravitational potential (Paper I), and
developed a method for fitting the orbit of the stream's progenitor
given the observed properties of the stream (Paper II).  
It is worth repeating here M31's
heliocentric radial velocity of $-300 \kms$ \citep{devauc91}.
Also, we assume the distance to M31 is $z_g = 784 \pm 24 \kpc$
\citep{stanek98}, which agrees very well with the distance of $z_g = 785
\pm 25 \kpc$ derived from the RGB maps by \citet{mcconnachie05}, and
implies that $1 \degree$ in angle corresponds to $13.5 \kpc$.

To construct a model of the progenitor, we first estimate its
orbit, using the methods similar to those in \citetalias{fardal06}.  
For the NE shelf we assume a central position on the sky of 
$\xi = 1.8 \degree$, $\eta = 0.7 \degree$.
We fit a progenitor trajectory to pass in the direction 
(not necessarily the exact location) of this point,
while still reproducing the observational
properties of the southern stream, obtained by
analytically ``stretching'' the trajectory to approximate the
phase-space location of the stream 
(\citepalias[see][for details]{fardal06}).
We also incorporate the radii and velocities of the 
\citet{merrett06} northeastern ``Stream'' PNe into the fit,
constraining the orbit to pass near their projected radii and
line-of-sight velocities. 

The M31 potential model here is a member of the family of models
described in \citetalias{geehan06}, and includes a Hernquist (1990) bulge,
an exponential disk, and a Navarro, Frenk, \& White (1996, henceforth
NFW) halo.  The specific parameters used here imply a
low-mass disk.  The results are not strongly sensitive to the disk
mass, though, since the rotation curve places strong constraints on
the overall potential.

\begin{table*}
\caption{Simulation Parameters}
\label{table.ics}
{
\begin{tabular}{llll}
\hline
Parameter & Symbol & Value & Comments \\
\hline
{\it Satellite structure} \\
\hline
Progenitor Mass               & $M_s$                 & $2.2 \times 10^9 \Msun$ & Plummer Sphere \\
Progenitor Radius             & $a_s$                 & $1.03 \kpc$ & \\
Central Surface Density       & $\Sigma_{0,s}$        & $6.57 \times 10^8 \Msun \kpc^{-2}$ & \\
Central Velocity Dispersion   & $\sigma_{0,s}$        & $39.1 \kms$ & 1-dimensional \\
Particle Mass                 & $m_p$                 & $1.68 \times 10^4 \msun$ &  \\
Particle Number               & $N_p$                 & 131072 & \\
\hline
{\it Satellite orbit} \\
\hline
Initial position$^\ast$       & $x_0$                 & $-34.75\kpc$ & \\
                              & $y_0$                 & $19.37 \kpc$ & \\
                              & $z_0$                 & $-13.99 \kpc$ & \\
Initial velocity$^\ast$       & $v_{x0}$                 & $67.34 \kms$ & \\
                              & $v_{y0}$                 & $-26.12 \kms$ & \\
                              & $v_{z0}$                 & $13.50 \kms$ & \\
Final time                    & $t_0$                 & $840 \Myr$ & \\
\hline
{\it M31 structure} \\
\hline
Total Bulge Mass              & $M_b$                 & $3.24 \times 10^{10} \msun$ & Hernquist Sphere \\
Total Disk Mass               & $M_d$                 & $3.66 \times 10^{10} \msun$ & Exponential Disk \\
Total Mass inside 125 kpc     & $M(<125\;{\rm kpc})$  & $7.3 \times 10^{11} \msun$ & \\
Virial Mass                   & $M_{200}^\ast$        & $8.8 \times 10^{11} \msun$ & \\
			      & $M_{100}^\ast$        & $9.5 \times 10^{11} \msun$ & \\
Bulge Scale Radius            & $R_b$                 & $0.61 \kpc$ & \\
Disk Scale Radius             & $R_d$                 & $5.40 \kpc$ & \\
Disk Scale Height             & $z_d$                 & $0.60 \kpc$ & \\
Halo Scale Radius             & $r_h$                 & $7.63 \kpc$ & \\  NFW profile
Virial Radius                 & $R_{200}^\dagger$        & $195 \kpc$ & \\
\vspace{3mm}
                              & $R_{100}^\dagger$        & $253 \kpc$ & \\
Disk Central Surface Density  & $\Sigma_0$            & $2.0 \times 10^8 \msun \kpc^{-2}$ & \\
Halo Density Parameter        & $\delta_c$            & $4.41 \times 10^5$ & 
 characteristic density relative to critical  \\
\vspace{3mm}
Halo Concentration Parameter &$C_{200}\equiv R_{200}/r_h$ & 25.5 & \\
\vspace{3mm}
Maximum Rotation Velocity     & $V_{c,max}$           & $259 \kms$ & \\
\hline
\multicolumn{4}{l}{$^{\ast}$ $x$, $y$, and $z$ here are 
coordinates in a system where $x$ points to the E and $y$ to 
the N in the plane of the sky, and $z$ points into the sky.} \\
\multicolumn{4}{l}{$^{\dagger}$ $M_\Delta$ is the mass enclosed 
within radius $R_\Delta$ such that the mean density inside is $\Delta\rho_c$, 
where $\Delta=100$ or $200$ (see \protect\citetalias{geehan06}).} \\
\end{tabular}
}
\end{table*}

Our approximate analytic method describes only the central path of the
debris in phase space.  To gain a more complete picture of the debris
pattern, we turn to $N$-body simulations.  We continue to use a fixed
potential to represent the influence of M31, but the progenitor is now
represented by self-gravitating particles.  We consider a purely
stellar system.  It is hard to put much dark matter in the progenitor
without either underproducing the amount of stellar mass seen in the
stream, or overproducing its observed width and velocity dispersion
\citepalias{fardal06}.  The progenitor is given a spherical Plummer-law 
density profile, with rough values of the initial mass and scale radius 
$M_s \sim 10^9 \msun$ and $a_s \sim 1 \kpc$ (later to be refined).
We initialize the particles of this hot spherical system
with the ZENO package of J. Barnes.
The particle velocities are assumed to be isotropic, and are set by
solving an Abel integral for the energy distribution rather than
assuming a Maxwellian \citep[see][]{binney87}, giving an initial
configuration very close to equilibrium.

Using the orbit from the fit above, the satellite is set in motion
inbound and slightly past apocenter, minimizing initial transients
from M31's tidal forces.  The formation of the progenitor and its
accretion into M31's halo are not modeled here.  Once we determine the
current state of the debris, we should be able to shed some light on
the progenitor's prior history, but for now it lies outside the scope
of our work.

We run the simulations with the multistepping tree code PKDGRAV
\citep{stadel01}.  We set the spline softening length to a small
fraction of the satellite's Plummer scale radius,
$\epsilon= 0.1 a_s \sim 100 \, \mbox{pc}$.  The gravity tree uses an
opening angle $\theta=0.8$, the node forces are expanded to
hexadecapole order, and the individual particle timesteps $\Delta t_i$
are related to the particle acceleration $a_i$ by the criterion
$\Delta t_i < \eta (\epsilon / a_i)^{1/2}$ with $\eta = 0.2$.

We ignore the effect of dynamical friction in both our orbital fitting
procedure and in the $N$-body simulations.  This is reasonable given that
we are not aiming at an exact fit of the observations,
since the mass of the progenitor satellite is small relative to the
total mass of M31 ($\tsim 10^{-2}$), and since the Coulomb logarithm 
is small as well
($\ln \Lambda \sim 3$).  Even so, the specific orbital energy of the
progenitor decreases somewhat during the run, since the energy
required to tidally disrupt and heat the satellite comes at the
expense of the satellite's orbital energy.
We ignore any perturbations from M32 or NGC 205 as well.  Their mass
is low enough that a significant deflection would require an unlikely
close passage, and their orbital parameters are not known well enough
for us to treat this possibility with any reliability.

To relate the numerical simulations to observable properties of the
stream, we assume a fixed $M/L$ ratio throughout the stellar
population, or more precisely a fixed ratio of stellar mass to RGB
star numbers, since it is the latter class of objects that are used to
trace the observed stream and related debris.  We also assume a fixed
ratio of stellar mass to PNe number.  \citet{merrett06} showed that
there is good agreement between the number counts of PNe and M31's
$r$-band light profile, which traces the old stars and thus the
stellar mass quite well.

Our initial orbital fit yielded promising results.  We then ran
several more simulations to choose the mass and size of the satellite,
essentially through trial and error, to match the length of the
southern stream and other observable features.  We also attempted to
optimize the orbital parameters further using $N$-body simulations,
but without clear further improvement.  The parameters of our final
simulation are given in Table~\ref{table.ics}.  We note that we have
certainly not explored our large parameter space carefully in this
search.  Also, it is not yet possible to make quantitative comparison
to some of the most relevant observational data such as the RGB
spatial distribution, due to thorny issues such as incompleteness
and contamination from both foreground Milky Way dwarf stars and 
background galaxies.
Hence, we make no claims that our present satellite orbital parameters
are optimal.  They should, however, suffice to demonstrate the basic
physical and observational properties we can plausibly expect from
this general scenario.

Our spherical Plummer model for the satellite with initial mass 
$M_s = 2.2 \times 10^9 \msun$ and scale radius $a_s = 1.03 \kpc$
produces a central velocity dispersion of $39 \kms$, consistent with
typical velocity dispersions for galaxies of this mass
\citep{dekel03}.  The observed stream stars appear to have a
moderately high metallicity, [Fe/H] $\approx -0.5$
\citep{mcconnachie03,raja06,kalirai06a}, and intermediate to old ages of 4--$11
\Gyr$ \citep{brown06,brown06b}.  Again comparing to \citet{dekel03}, 
this observed
metallicity is in very good agreement with the assumed progenitor
mass.  If we assume a $V$-band $M/L$ ratio of 5 in the progenitor,
appropriate for a baryon-dominated galaxy with an old stellar population
and a Salpeter initial mass function, 
the central surface brightness is $\Sigma_0 \approx 1.3 \times 10^8 \,
L_{\sun,V} \kpc^{-2}$.  This is a factor of 3 below the mean
regression line of Dekel \& Woo, but is still consistent with the
scatter about this line.  The Plummer law is not particularly
concentrated towards the center.  If we had adopted an $R^{1/4}$ 
law for the satellite, the central surface brightness would have 
been about 300 times larger, for the same total luminosity and
projected half-mass radius.

We run the satellite model with PKDGRAV for 2 Gyr.
The first pericentric passage occurs at 0.17 Gyr, giving rise to the 
southern stream.  The first apocenter occurs at 0.44 Gyr and the
second pericenter at 0.70 Gyr.  We select
the output that most closely resembles the spatial and kinematic 
properties of the stream and shelf debris.  This output occurs at 
0.84 Gyr past the start of the simulation
(Table~\ref{table.ics}).  At this point the progenitor
has been completely disrupted and its central stars lie in the 
location of the observed NE shelf.  We first tried runs with a small
number of particles before proceeding to the final run;
we did not find any significant differences resulting from 
particle resolution.

In several of the plots below, we will use only the simulated satellite 
particles for clarity of presentation.  However, when comparing to the
observations, it is essential to have an idea of the characteristics
of the main stellar body of M31, which in projection is coincident
with much of the
satellite debris.  Both \citet{ferguson05} and \citet{brown06} found
that at 10--20 kpc from M31's center, the stellar populations of the 
spheroid and of the stream were quite similar.  On the other hand, 
an outer halo component has recently been discovered
in M31 \citep{raja05,irwin05,kalirai06b,chapman06}; incorporating these new
observations, the mean metallicity of the smooth M31 spheroid is seen 
to decreases radially outward down to
[Fe/H] $\approx -1.4$ at $R \gtrsim 60 \kpc$.
In contrast, the stream's mean metallicity is more or less constant 
along its length, out to similar projected radii.  So depending on
location, the characteristics of the stellar population may or may not
be useful for separating the contributions of the stream debris and
M31; in some regions kinematics may be the only guide.

Therefore, we construct an $N$-body particle realization of M31
itself to mimic the observed stellar components of M31, and combine it
with the satellite debris to predict the full spatial and velocity
pattern.  We only add this particle realization at the end, rather 
than evolving it dynamically with the $N$-body code.
The model in \citetalias{geehan06} predated the discovery of
the metal-poor halo component just mentioned; it is
important to include this in the model to estimate the M31
contribution at large radii.  We model the halo stars by assuming that
a small fraction of the NFW halo mass 
is composed of stars rather than dark matter,
equivalent to using an NFW profile normalization $\delta_{c,\ast} = 350$.  
To fit the profile better, we also truncate the
Hernquist bulge past $r_{tr,b} = 10 \kpc$ since it would otherwise overproduce
the surface brightness near 20 kpc.  We use the default truncation scheme in 
ZENO, which cuts off $dM(r)/dr$ as an exponential past the truncation radius.
(We neglect this truncation in the orbit calculation, since it changes
the mass within 125 kpc by only 0.2\%.)
With these modifications, the composite stellar profile exhibits a 
change in slope at $\tsim 30 \kpc$ similar to that in the observed profile
\citep{raja05,irwin05,kalirai06b}.  

For our $N$-body realization of M31, we use the same particle mass
as for the progenitor particles, and again use the same constant
scaling of RGB or PNe number to stellar mass.
Given that there are slight but observable 
stellar population differences between M31's giant stream and spheroid 
\citep{ferguson05,brown06,kalirai06a,kalirai06b,brown06b},
our constant scaling assumption is a slight oversimplification.
Moreover, there are significant observed deviations from our simple
model in the form of spiral arms, warping of the disk, and elongation
of the bulge and inner disk.  Our model should thus be taken as an
illustrative proxy for the M31 stellar component, rather than a
precise description.

\subsection{Shelf morphology}
\label{sec.morphology}

\begin{figure*}
\includegraphics[width=16cm]{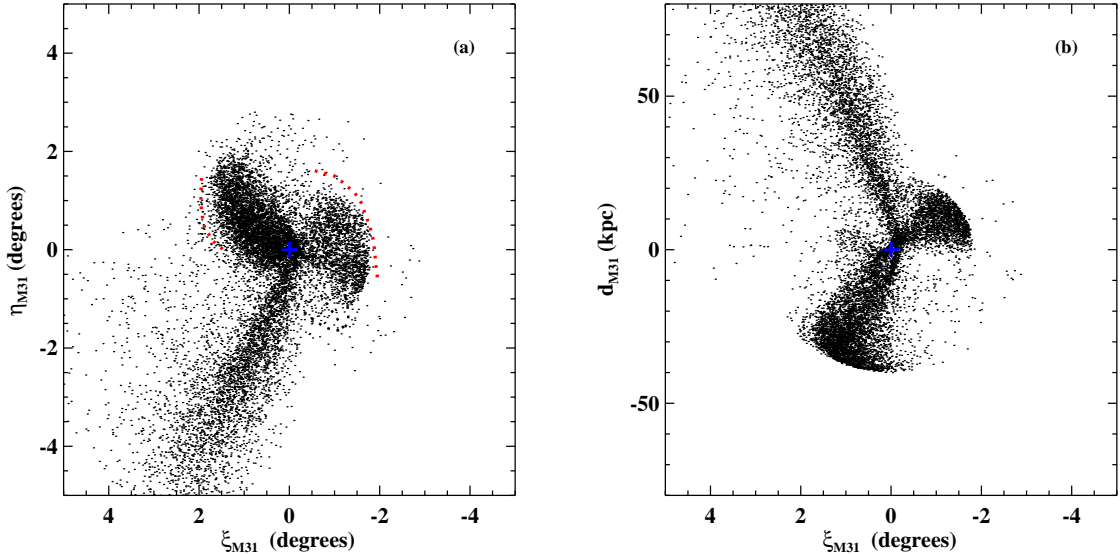}
\caption{
\label{fig.ppos}
The spatial distribution of satellite debris particles in the
simulation, centered on M31.  {\it Panel a:} sky projection in standard
coordinates. This can be compared to the observed star-count map in Figure
1a.  The red dotted lines show the observed shelf edges from that
figure.  For clarity, a randomly chosen subset of particles is shown
in this and most subsequent figures.  The particles of our static M31 model
are also omitted; the disk component of that model would dominate much
of the NE shelf but only a small portion of the W shelf.  {\it Panel b}:
the view of the debris looking down the north axis in Panel (a).  
``Depth'' $d_{M31}$ increases with distance from Earth.
The scale on that axis matches the angular
scale at M31's mean distance of $\approx 780$~kpc.  The NE shelf is directed
mainly toward us, while the W shelf is oriented away from us
close to the plane of the sky.}
\end{figure*}

We begin the comparison to observations by examining the pattern on
the sky of the NE and W shelves.  The spatial distribution of the
particles in our simulation is shown in Figure~\ref{fig.ppos}.  The
debris resembles the observed shelves (shown in Figure~1) on both the
E and W sides, with a much larger surface density on the E side.  The
edges of these shelves are at similar, but in some places slightly smaller 
radii to the observed edges.  The shelves also cover about the right
azimuthal range, though the faintness of the features and interference from
M31 itself make it difficult to make a precise comparison

We note that the projected radius of the NE shelf from M31's center,
as shown in Figure~\ref{fig.sky} and given in Table~\ref{table.edges},
appears to vary with azimuth, instead of forming a circular arc.  This
may be a significant clue to the nature of the shelf.  In contrast,
the edge of the W shelf appears more circular.  The simulations appear
to reproduce the behavior of projected radius as a function of azimuth
for both shelves.

We next turn to the surface densities. In the star-count map of 
\citet{irwin05}, the brightness of the NE shelf
appears similar to the denser, innermost distinct parts of the S
stream (at $R_{proj} \approx 1.6 \degree$), 
while the brightness of the W shelf appears more similar to
the more tenuous middle parts of the S stream
(at $R_{proj} \approx 2.5 \degree$).  To obtain a crude
estimate of the surface densities in the latter regions, we first construct a
simple estimate for the southern stream surface brightness using the
transverse and longitudinal profiles in
\citet{mcconnachie03}. We normalize these profiles by using the stream's
$V$ magnitude of $M_V = -14.0$ or 
$V$ luminosity of $3.4 \times 10^7 L_{\sun,V}$ in the range 
$15 \kpc \ltrsim R_{proj} \ltrsim 40 \kpc$
from M31's center \citep{ibata01}. Assuming a stellar 
$M/L_V \approx 5$ as before, this gives a stream stellar mass 
within this radial range of $\sim 2 \times 10^8 \msun$.
Computing the surface density of the resulting mass model in
the areas mentioned above, 
we can then estimate that the NE and W shelf surface densities are
very roughly
$\tsim 1.0 \times 10^6 \msun \kpc^{-2}$ and $\tsim 8 \times 10^5
\msun \kpc^{-2}$ respectively.   We arbitrarily define the inner radial 
boundary of both shelves to be 60\% of their outer radii (listed in 
Table~\ref{table.edges}) in order to
isolate the regions in which the shelves have the highest contrast
against the smooth M31 components;  the NE shelf has an
area $150 \kpc^2$ and the W shelf has an area $200 \kpc^2$.  
This then
implies masses of $\tsim 1.5 \times 10^8 \msun$ within both shelf
boundaries.  We have not attempted here to subtract the background
density of M31 disk and spheroid stars, so these surface densities and 
masses are only lower limits.  Also, these estimates only
apply to the mass inside our assigned boundary; if the shelf mass
distribution continues radially inwards onto the face of M31, 
the total mass could be
much larger.  Given all the uncertainties, these are clearly 
only order-of-magnitude estimates, but even so should be good enough 
to check the plausibility of the model.

We compute the mass of the satellite debris particles 
in the simulation that lie within the 
boundaries of the NE and W shelf regions as defined above, and find
$1.2 \times 10^8 \Msun$ for the NE shelf and 
$1.4 \times 10^8 \Msun$ for the W shelf (although these regions
clearly do not select all the mass in the NE and W radial lobes).
These imply surface stellar mass densities of 
$8 \times 10^5 \Msun \kpc^{-2}$ for the NE shelf and 
$7 \times 10^5 \Msun \kpc^{-2}$ for the W shelf.
The former surface density in particular is a slight underestimate
because it does not extend all the way to the observed shelf edge.
Inclusion of the particles in the static M31 model raises these
surface densities to 
$1.1 \times 10^6 \Msun \kpc^{-2}$ for the NE shelf and 
$1.0 \times 10^6 \Msun \kpc^{-2}$ for the W shelf.
Comparison of these values with the rough observational estimates 
above shows that our model has passed another test.

M31 is close enough that the distribution in three spatial dimensions is
measurable.  \citet{ferguson05} used color-magnitude diagrams with
HST/ACS to measure the brightness of red clump stars in the stream,
21 kpc in projection from the center of M31, and in the NE shelf, 
at $\xi = 1.26 \degree$, $\eta = 0.24 \degree$.  They found that 
stars in the ``NE shelf'' fields indeed lay closer to us than
those in the ``Giant Stream'' fields, by a factor $1.07 \pm 0.01$.
With the ``depth'' $d$ (defined here as the distance behind M31) 
in the stream field estimated to be $(50 \pm 20) \kpc$ by
interpolating the results of \citet{mcconnachie03}, this would 
put the shelf at depth $d_{\rm F05} = (0 \pm 20) \kpc$.  However,
\citet{brown06} measured the red clump brightnesses at a nearly
identical point in the stream and in a minor-axis, spheroid-dominated
field, and found the stream there was only $(11 \pm 5) \kpc$ further
than the spheroid position, which is likely at 
$d \approx 0$.\footnote{Flattening of the spheroid could introduce
a systematic offset to this estimate, to larger distances if the
spheroid is aligned with M31's disk.  
However, the consistency to within $\ltrsim 0.03$~mag 
of the spheroid fields of \citet{brown06}, at $R_{proj} = 12 \kpc$, 
and \citet{ferguson05}, further out at $R_{proj} = 20 \kpc$, 
suggests that the Brown \ea field is shifted
by $\ltrsim 0.03$~mag or $\ltrsim 12 \kpc$ with respect to M31's nucleus.}
From this estimate we can infer the shelf is at $d \approx -40 \kpc$, 
in front of M31.

Our model implies the material forming the NE shelf lies on average
well in front of the center of M31, and hence even farther in front of
the southern stream (see Figure~\ref{fig.ppos}).  In the field of
\citet{ferguson05}, the median depth of the simulation particles is
$d_{\rm F05} = -31 \kpc$, in agreement with the estimate based on
\citet{brown06}, though the distribution is highly skewed with a small 
second peak at $\approx -5 \kpc$.  The model thus appears to pass the test 
of the observed distances as well.

\subsection{Velocity measurements}

\begin{figure*}
\includegraphics[width=16cm]{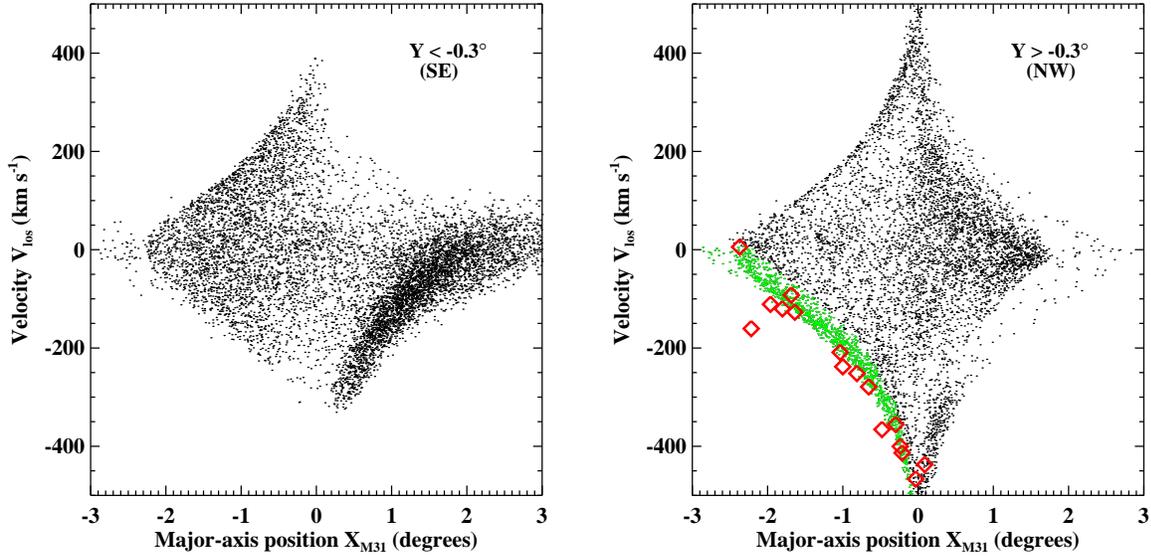}
\caption{
\label{fig.xv}
Plots of the velocities of satellite debris particles relative to M31 versus
major axis position $X$, split on the basis of minor-axis position $Y$
into two panels.  $X$ increases toward the SW, and $Y$ toward the NW.
Left panel: southeast portion ($Y < -0.3 \degree$).  
The dense concentration of points at
the lower right edge is the southern stream.  
Right panel: central and northwest portion ($Y > -0.3 \degree$).  
The PNe in M31 notated by \citet{merrett06} as ``Stream'' or
``Stream?'' are marked by red diamonds.  The dense
concentration of $N$-body particles within $60 \kms$ of
the lower left edge are marked in green; we regard these particles
as an analogue to the Merrett \ea ``PNe stream''.  }
\end{figure*}

\begin{figure*}
\includegraphics[width=16cm]{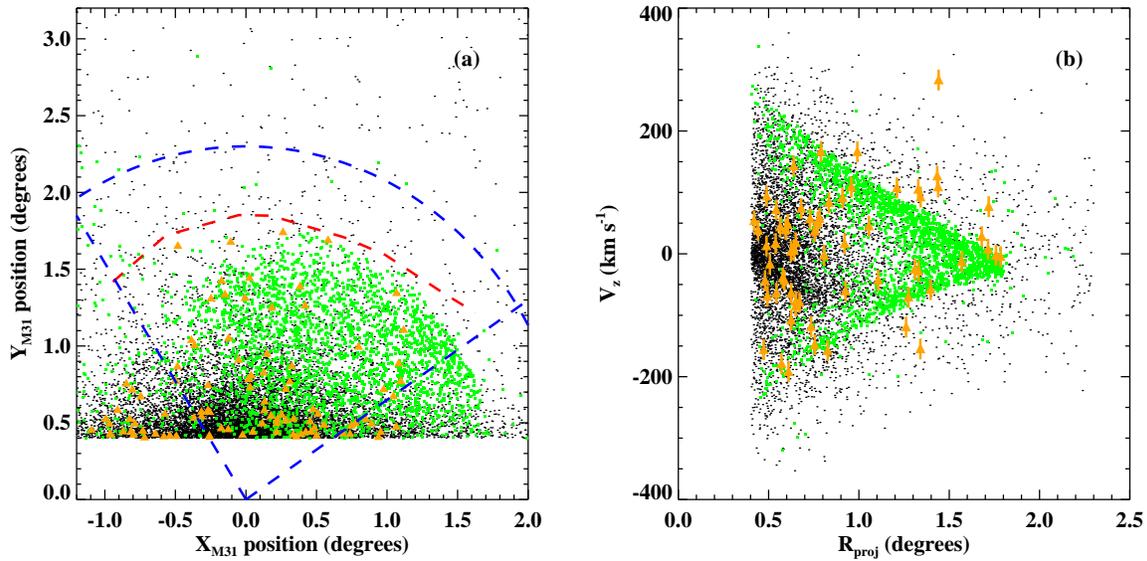}
\caption{
\label{fig.wpne}
Plots of $N$-body particles and observed PNe in the W shelf region.
Objects with minor axis coordinate $Y < +0.25 \degree$ are omitted.
{\it Panel a:} sky position in M31-axis coordinates $X$ and $Y$.
Green points show satellite debris particles, while
black points show particles from the model of M31 itself.
Orange triangles show observed PNe from \citet{merrett06}, 
excluding those noted there as associated with NGC 205.
The red dashed line shows the observed W shelf boundary
from Figure~\protect\ref{fig.sky} and Table~\protect\ref{table.edges}.
The sky coverage in the PNe survey extends roughly out
to the outer blue dashed arc.   Only objects shown within the 
blue dashed lines are selected for the next panel.  
{\it Panel b:} projected radius versus line-of-sight velocity.
The symbols are the same as before.
The PNe velocities have a measurement error of $17 \kms$.
Note the group of PNe close to the positions and velocities
of the simulated shelf particles; this association is tested
statistically in the text.
}
\end{figure*}

The overall velocity pattern in our simulation is shown in
Figure~\ref{fig.xv}.  We will discuss the physical origin of the
features in this plot later, in \S~4.3.  Here we only wish to compare
the simulated velocities to the observed ones to see if they agree in
their main respects.
At present, there are three large-scale surveys that can constrain the
overall velocity pattern of the observed debris: the wide-area survey
of PNe by \citet[][2006]{merrett03} and two ongoing surveys of RGB
velocities in selected widely spaced fields 
\citep{ibata04,ibata05,chapman06,raja06,kalirai06a,kalirai06b,gilbert06}.

The PNe detected by Merrett \ea include a component that moves
opposite to the disk rotation on the NE side, with a narrow velocity
width, suggestive of a continuation of the southern stream.  
The PNe indicated by Merrett \ea as ``Stream'' or ``Stream?''
are shown as triangles in the right panel of Figure~\ref{fig.xv}.  
This apparent stream is unlikely to be produced by the usual
stellar components of M31, so it is strongly indicative of some kind
of tidal debris.  
The simulation produces a similar ``stream'' of material, due to a set
of stars in the N part of the NE shelf which have similar velocities;
like the observed PNe stream, these overlap the major axis of M31,
as shown in Figure~\ref{fig.fan} below.

The Merrett \ea survey extends into an irregular area in the halo
region of M31.  In this region, Merrett \ea used wide-field images to
optimize the choice of fields for closer study.  These fields
are sparse, so the survey coverage may change with position;
we will ignore this potential complication.  
Figure~\ref{fig.wpne}a shows the survey results
in the area of the W shelf; the survey coverage 
within the W shelf boundary (dashed red line) and between the
dashed blue radial lines is almost complete, while it is
sparser but not zero outside the W shelf boundary.  We exclude
the PNe that Merrett \ea find to be associated with NGC 205 on the
basis of their positions and velocities.  The same plot shows
the simulation particles, including both the satellite
debris (in green) and the particles from the static M31 model (in
black).  Although the number of PNe is limited, it seems that the
density of PNe does not fall off all the way towards the W shelf;
beyond $R \sim 0.7 \degree$ it remains roughly constant,
suggesting that the PNe also trace the shelf.  We plot the velocities
of the PNe and simulation particles within the dashed selection boundary 
in Figure~\ref{fig.wpne}b.  The satellite debris in the simulation 
forms a triangular shape in this plot.  Interestingly, the PNe at
large $R_{proj}$ mostly fall near the boundary of the simulated satellite 
debris; it appears that increasing the shelf radius or particle
velocities slightly in the simulation would result in even better
agreement.

To test this apparent agreement quantitatively, we
compare two hypotheses: that the distribution matches the pure M31
particle distribution, or that it matches the sum of the M31 and
satellite particles.  We bin the particles in a grid covering the
relevant area in projected radius and velocity, after convolving their
velocities with a Gaussian distribution of $17 \kms$, which is the
random error in the PNe velocities.  We then normalize the
distributions, both including and excluding the satellite debris
particles, to unity.  The product of these binned distributions at the
locations of the PNe then gives the likelihood function of these two
hypotheses.  We find that the likelihood ratio is about 180, meaning
that the model with satellite debris is 180 times more likely than the
model that excludes it, assuming equal prior probabilities.
Furthermore, if as suggested by Figure~\ref{fig.wpne},
we add $30 \kms$ to the velocities of the satellite particles as 
might be induced by a change to the overall angular momentum,
the likelihood ratio increases to 4500.  Thus these tests appear to
show that some of these PNe are bona-fide members of the W shelf, and
confirm its overall kinematic structure.


The remaining M31 radial velocity surveys use RGB stars.  Previous analyses of
substructure in these surveys mostly focus on the giant southern stream,
the extended disk-like component \citep{ibata04,ibata05,raja06}, 
and the loop-like features near NGC 205 \citep{mcconnachie04},
but there are some hints of other structures.
The stream of ``counterrotating'' PNe may have a counterpart in
the inner disk RGB velocity measurements of 
\citet[][their Figure 9]{ibata05}, though
this feature is not discussed in that paper.  
\citet{chapman06} mention kinematic substructure in the
NE that they attribute to a continuation of the southern stream,
but the exact location of the responsible fields is not clear.

\begin{figure}
\includegraphics[width=8cm]{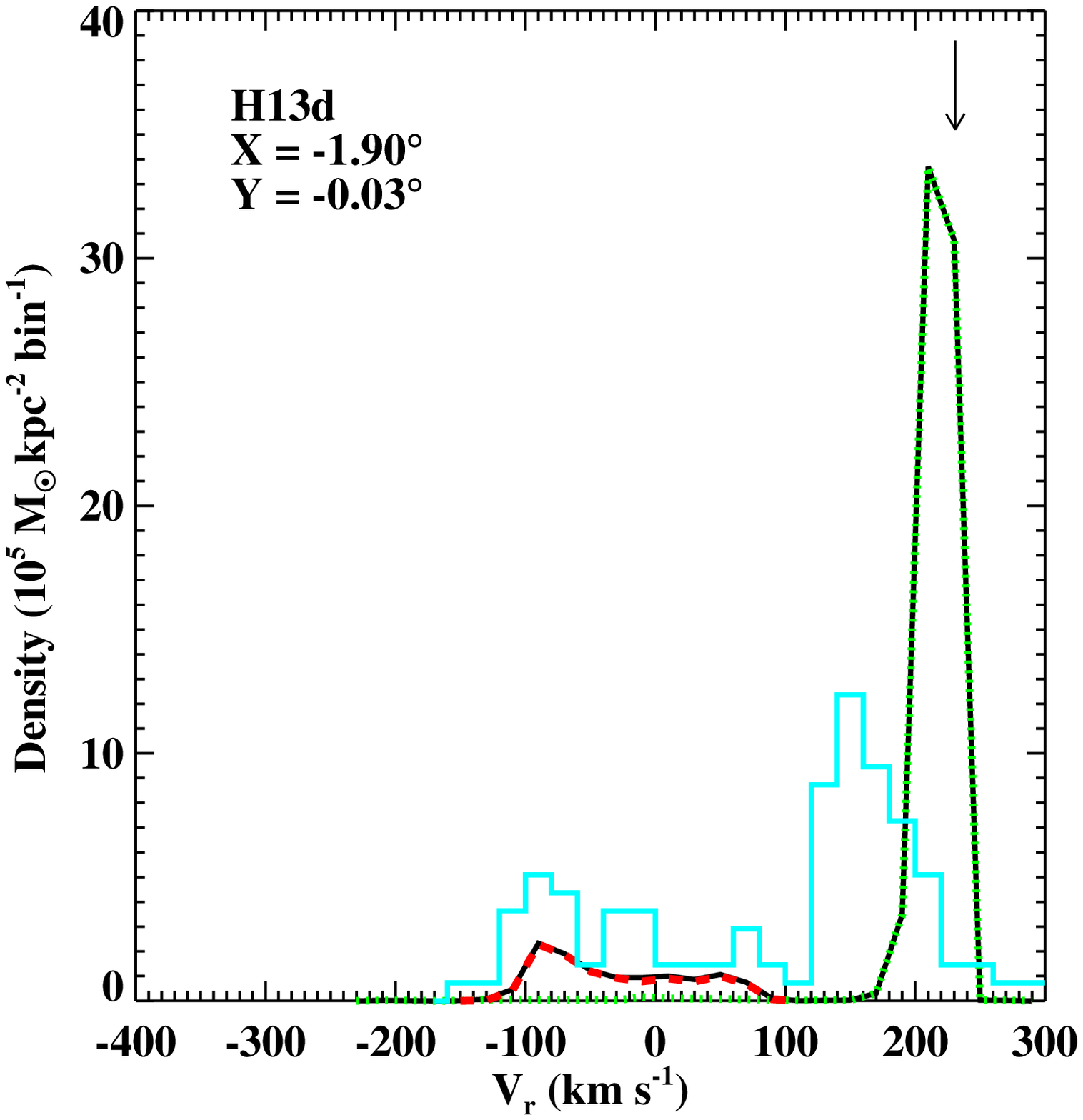}
\caption{
\label{fig.vdiskfield}
The velocity distribution relative to M31 in the field H13d of 
\citet{reitzel06} (also discussed in \citealp{kalirai06a}).
The field used to select the simulation particles is taken to
be a square $0.3 \degree$ on a side, to ensure a large sample of
particles.  Simulation results using the satellite debris alone are
shown by a red dashed line; results using the M31 galaxy model alone
are shown by a green dotted line; and results using both satellite and
M31 particles are shown by a thick solid line.  The distribution of
observed RGB star velocities is
shown with a cyan histogram; this is normalized to have the same
total density as the full simulation histogram.  The disk velocity
from the empirical model in \citet{ibata05} is shown by an arrow.
The observations show at least a qualitative match to both the
prograde (disk) and retrograde (satellite debris) components in the
simulation, as discussed further in the text.
}
\end{figure}

The RGB surveys obtain a much higher density of target objects than
the PNe survey, albeit with a much sparser spatial coverage.
Thus it is worth plotting the velocity distribution in each field as a
separate histogram.  We begin with the field H13d of 
\citet{kalirai06a} and \citet{reitzel06}. 
The observational and simulation results are 
shown in Figure~\ref{fig.vdiskfield}a.  We include three
curves that show the results of the simulation for the satellite
debris alone, for the M31 model stars alone, and for the combination
of the two.  Here we {\em assume} the total mass density in the
observations matches the total found in our simulation, and normalize
the observational histogram accordingly.  This method allows us to
avoid the complex issues of the selection and detection
efficiencies in the spectroscopic survey.  

A substantial population at negative velocities, opposite to the sense
of disk rotation, is evident in the observational histogram.  This
component is not visible at all in the galaxy model, but it matches
the velocity distribution of the satellite debris quite well, even if
the overall normalization is somewhat off.  Another interesting feature
is the difference in mean velocity, velocity dispersion, and possibly
total mass between the observed and predicted disk components; 
we will return to this point below.

\begin{figure*}
\includegraphics[width=16cm]{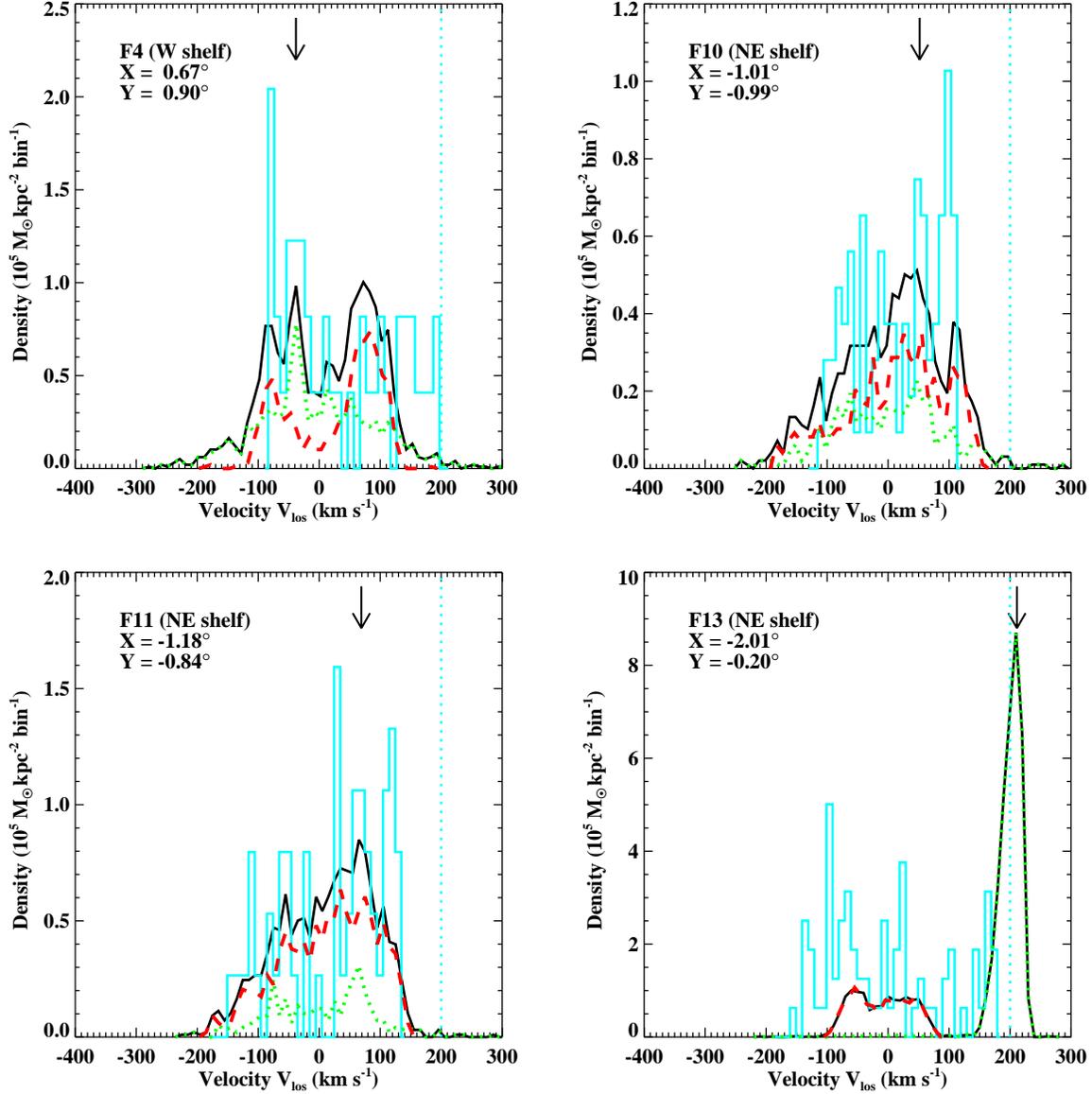}
\caption{
\label{fig.vhalofield}
Same as Figure \protect\ref{fig.vdiskfield} 
for the four spectroscopic fields observed for RGB star velocities 
by \citet{ibata05} that lie
in the shelf regions.  Otherwise the caption to the previous figure
applies here.  The observational histograms are generated by taking the
histogram of ``disk lag velocity'' in \citet{ibata05} and shifting by
the velocity in their model of the disk at the center of the field.
This neglects the slight variation of the disk velocity across the
field.  In these observational histograms, stars with positive 
velocities greater than $200 \kms$ relative to M31 are removed 
to reduce Galactic contamination, as indicated by the dotted line.
The observations show evidence for satellite debris in at least
Panels (a) and (d), as discussed in the text.
}
\end{figure*}

The spectroscopic survey of \citet{ibata05} also includes 16 fields in
the outer disk.  In at least 7 of these fields, the velocity
distribution shows a strong, narrow component near the expected
velocity of the disk, which forms the focus of that paper.  None of
the other fields is dominated by the disk component, although a disk
component emerges when the fields are stacked together.  As shown in 
Figure~\ref{fig.sky}, four of these
latter fields lie in the regions possibly dominated by debris
associated with the NE and W shelves, namely F10, F11, and F13 in the
NE shelf, and F4 in the W shelf.  The observed
velocity distributions in these fields are shown in
Figure~\ref{fig.vhalofield}, together with the prediction from our
simulation.  One complicating factor is that Ibata \ea remove all
stars with velocities greater than $200 \kms$ (or $-100 \kms$
heliocentric), to reduce contamination from MW dwarf stars.  They also
plot histograms of the velocity relative to their disk velocity model;
we have transformed these to M31-centric velocity using their model
velocity in the center of their field, but this correction neglects the
slight variation of this model velocity across the field.  The shape of
the histogram is thus not exact, especially near the 
velocity limit of $200 \kms$ (shown with a dotted line) 

In F4, F10, F11, and F13, stars are detected in a broad velocity
distribution up to $|v| \sim 200 \kms$ relative to M31, as shown by
the light cyan histograms in Figure~\ref{fig.vhalofield}.
A glance at Figure~\ref{fig.vhalofield} shows that in fields F10 and
F11, the predicted satellite and static M31 velocity distributions are 
actually fairly similar.  Thus, the velocity distribution in these two
fields cannot serve as evidence for or against our scenario without a
deeper understanding of the slight differences.  The histograms from
the simulations suggest that the satellite debris dominates the
distribution in both fields; differences in the stellar populations
might serve to test this result, though this also could be difficult 
as these differences are quite subtle at this radius 
\citep{ferguson05,brown06,brown06b}.

In field F4, the strongly bimodal satellite distribution biased to
positive velocities improves the agreement between the simulated and
observed velocities, though the observed distribution extends to even
higher velocities than shown by the simulated satellite debris.  It is
worth noting that the problem of foreground MW dwarf contamination
increases with the velocity.  The observed and simulated
results in field F13 are similar to those in H13d above; again there
is a population of observed stars at negative velocities, which are
explained fairly well with the satellite debris.  These two fields are
thus encouraging for the debris model.

Field F13 also resembles Field H13d in that there is a strong
predicted galaxy disk peak that is not seen in the observations.  Once
again there may be a smaller disk peak shifted to lower velocity by
40--$60 \kms$.  The velocity cut at $200 \kms$ imposed by Ibata \ea to
select M31 stars could be
responsible for at least part of the difference, as it would remove
many (though not all) disk stars in our $N$-body model for M31.
Moreover, the warp of M31's disk, which is quite strong this far along
the NE major axis, may cause the disk contribution to be substantially
lower in field F13 than in H13d (F13 being slightly off the major axis
in the direction away from the warp).  However, the H~I in this region
shows a peak at $220 \kms$ \citep{newton77}, exactly as predicted by
the disk model, even though the H~I disk itself is warped.  A further
possibility is that unlike the H~I, the stellar velocities are
perturbed from circular orbits.  The extended disk found by
\citet{ibata05} shows large velocity and density fluctuations relative
to a smooth disk, and the results in fields H13d and F13 may be a
further example of these fluctuations.  A difference in H~I and stellar
velocities might be expected if the extended stellar disk resulted from
an accreting satellite, as in the model of \citep{jorge06}.

In summary, it appears that the debris model studied here explains a
number of irregularities seen in the spatial and velocity distribution
of the stars and PNe in and around M31.  The strongest pieces of
evidence for the model are the shape of the NE shelf on the sky, the
apparent stream of counterrotating PNe and RGB stars seen along the NE
major axis, and the agreement of the W shelf in the model with the
spatial pattern seen in RGB stars and the kinematic distribution of
PNe in the same area.  The measured distance to the NE shelf stars is
also consistent with our model.  The fields observed for RGB
velocities to date are not in regions that are optimal for testing the
model, even though the existing fields are encouraging for the debris
model, so we can anticipate great improvement in the observational
constraints on this model.  We now examine how future observations can
test the model for the debris pattern, how we can understand its
origin physically, and how we can use the debris to better understand
M31 and its environment.

\section{PHYSICAL INTERPRETATION AND DIAGNOSTICS}

\subsection{Model for collisionless shell systems}

The NE and W shelves in our model appear similar to the ``shells''
observed around a significant fraction of elliptical and S0 galaxies.  
One mechanism for forming shells involves the disruption and folding
of a thin, cold disk \citep{quinn84,hernquist88,barnes92}.  This scenario
may well occur in nature, but it would tend to produce sharp, asymmetric
features with no clear orientation around the galaxy center (for
example, see Figure~3 of \citealp{quinn84}), and hence be a poor fit to the
best-defined examples of shell patterns in ellipticals, as well as the
tangentially oriented shelves discussed here. 
A more likely creation mechanism in our case is the accretion of a
satellite on a nearly radial trajectory
\citep[e.g.,][]{schweizer80,hernquist88,barnes92}.  Once the satellite is
disrupted in whole or in part by its first close passage, the
correlation between orbital energy and orbital period sorts the newly
unbound stars in order of energy into a dynamically cold stream, 
with the lowest-energy stars executing the smallest and fastest orbits.  
The cooling of the stream means the progenitor 
need not itself be dynamically cold to produce a coherent shell, 
though this is possible as well.  Subsequent passages
near the center of the galaxy, combined with the dispersion in the
angular momenta of the stars, disperse the stream stars in a variety
of directions.  The energetic sorting ensures that they nevertheless
reach turnaround at a similar radius at any given time.  This ongoing
phase-wrapping process creates a series of caustic or ``fold''
features located on spherical shells, whose radii decrease along the
orbit.  

This scenario is much like our current understanding of the creation
of M31's southern stream (\citealp{ibata01,ibata04,font06}; Paper~I).
Hence shell-like features are expected in M31 as long as the debris
extends far enough ahead of the stream.  The previous section provides
evidence that this is indeed the case.  In this subsection we discuss
in more depth the physical properties of radial shells.

\subsubsection{Stellar orbits in shell systems}
\label{sec.orbits}
To provide a concrete and analytically tractable example, let us make
a number of approximations.  First, we assume a power-law potential
for the primary galaxy with $\Phi(r) \propto -r^k$.  In \citetalias{fardal06}, 
we found this was a reasonable approximation to the actual potential 
of M31 in the radial range 3--$100 \kpc$, with $k \approx -0.4$.

When the satellite collides with the primary galaxy, the tidal forces
alter the energies of stars depending on their positions and orbits
within the satellite, much like the gravitational slingshot of a
planetary probe.  In turn, the dispersion in their specific energies 
is related to the progenitor satellite's mass $M_s$ and pericenter $r_p$, 
as found in \citet{johnston98} and confimed by the simulations
in \citetalias{fardal06}.  Using the latter simulations we estimate
\begin{equation}
\label{eqn.edispersion}
\sigma_E \approx 0.7 
   V_c^2(r_p) \left( \frac{G M_s}{r_p V_c^2(r_p)} \right)^{1/3} \; ,
\end{equation}
which we also found to be valid for our run in this paper.  Here
$V_c(r) = (G M(r)/r)^{1/2}$ is the circular velocity.  The dispersion
then increases with decreasing $r_p$.  The combined requirements of a
large dispersion in the energy and large angular deflections at
pericenter explains why shell systems are most easily modeled using
satellites with small $r_p$, i.e., on nearly radial orbits.  We should
note that we have not tested the validity of Equation~\ref{eqn.edispersion}
outside the regime of nearly radial orbits and an M31-like potential.

Next, we assume the newly liberated stars travel on 
purely radial orbits, i.e., we ignore the angular momentum.
This is sufficiently accurate in our scenario, where the unperturbed orbit 
of our progenitor has a pericenter 45 kpc and apocenter 2 kpc.
The stars then take on
orbits that are described by the energy alone, or equivalently their
orbital periods $t_r$.  Let the turnaround radius of the progenitor be
$r_{t0}$ and the circular velocity there be $V_{t0}$.  Then $V_c(r) =
V_{t0} (r/r_{t0})^{k/2}$.

For an arbitrary star, let us define a time scaling factor $\tau$
proportional to the orbital period of the star, $\tau = t_r / t_{r0}$.
The self-similar nature of the orbit family leads to the scalings
$r_t \propto \tau^{2/(2-k)}$, $V_t \propto \tau^{k/(2-k)}$,
and $E \propto \tau^{2k/(2-k)}$.  
Stars with lower energies move faster, as long as $k < 0$.  The radius
and velocity of a star with a given $\tau$ can be obtained from the
orbital trajectory of the progenitor using the relations
\begin{eqnarray}
r(t) & = & \tau^{2/(2-k)} r_0(t/\tau) \; , \\
v(t) & = & \tau^{k/(2-k)} v_0(t/\tau) \; .
\end{eqnarray}
Here we are treating the progenitor as a test particle, i.e., we
ignore any change in the progenitor's orbit from tidal disruption and
dynamical friction.  
The exact orbital period of the star is 
\begin{equation}
\label{eqn.period}
t_r = \sqrt{\frac{2}{(-k)}} \, B[1/2,(2-k)/(-2k)] \frac{r_t}{V_t} 
\end{equation}
where $B$ is the beta function.

The turnaround radius of a shell is found for $dr/d\tau = 0$,
which results in the requirement
\begin{equation}
\frac{t}{\tau} v_0(t/\tau) - \frac{2}{2-k} r_0(t/\tau) = 0 \; ;
\end{equation}
solving for the roots of this equation yields a series 
of spherical shells at fixed orbital phases.
The radius of the shell at a given orbital phase is then 
$r_s = \tau^{2/(2-k)} r_0(t/\tau) \propto t^{2/(2-k)}$.  The shell
moves outwards with time, as stars with higher and higher
energies arrive at the shell orbital phase.  
The ratio of the radii of two different shells is constant with time.
These results are specific to the power-law potential, but also
give a sense of the shell properties in the more general case.

\subsubsection{Idealized surface density}

\begin{figure}
\includegraphics[width=8cm]{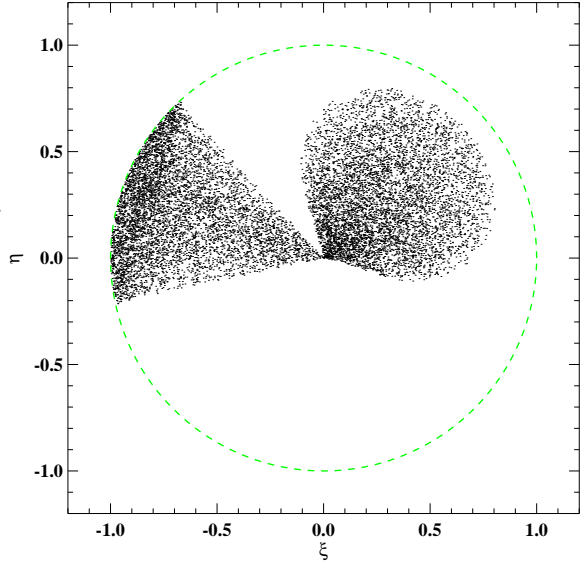}
\caption{
\label{fig.spherecartoon}
Idealized illustration of partial shells in a spherical potential.
Two shells are shown, which are identical except in their orientation.
Each shell fills a cone with an opening angle of $30 \degree$; the
radial distribution assumes a steady stream in a power-law potential.
The leftmost shell is tangent to the line of sight, and thus its outer
boundary on the sky is defined by the shell radius.  Note the dark
region where the shell's ``end cap'' is visible.  The rightmost shell
is not tangent to the line of sight, and as a result has a less
distinct edge, and a boundary that varies in projected radius.}
\end{figure}

To understand the sky pattern of a shell system, we must realize that
the stars in a given shell may cover only a small fraction of
the sphere, particularly early in its history as is the case for our
simulation (cf.\ Figure~\ref{fig.ppos}).  The observed
properties of the shell depend on whether or not the shell surface
lies tangent to the line of sight anywhere.  These two cases are
illustrated in Figure~\ref{fig.spherecartoon}.  

If the shell surface is tangent to the line of sight, as in the leftmost
shell in the figure, then the caustic
feature at turnaround is visible in projection as a circular boundary
of the shell material.  Let us here neglect locally the 
variation of the energy and mass flow rate along the stream.
All stars in such a stream share the
same shell radius $r_s$.  Near this shell radius, the velocity of a
star as a function of time $t$ is given by 
$dr/dt \approx -r_s^{-1} V_c^2(r_s) (t-t_s) = 
-\sqrt{2} V_c(r_s) [(r_s - r)/{r_s}]^{1/2}$,
where $r(t_s) = r_s$ and $V_c$ is the circular velocity.
Combining the inflowing and outflowing stars, we use this
to derive the mass density of stars as a function of radius:
\begin{displaymath}
\rho(r) \approx \frac{2}{4 \pi r^2 |dr/dt|} \frac{dM}{dt}\; \\
      \approx \frac{1}{2^{3/2} \pi \, f_\Omega \, r^2 \, V_c(r_s)} \,
         \left( \frac{r_s - r}{r_s} \right)^{-1/2} \, \frac{dM}{dt} \, .
\end{displaymath}
Here $dM/dt$ is the rate at which mass flows through the stream
(for example, the amount of mass that reaches apocenter per unit time).
$f_\Omega$ is the fraction of the total $4\pi$~sr of the
sphere that the shell covers; put another way, it specifies the ratio
of the local mass flow rate per unit solid angle to its average
over the sphere.

When integrated over constant {\em projected} radius, this gives the
result that the mass surface density $\Sigma$
approaches a constant just inside the shell radius,
\begin{equation}
\label{eqn.edgesurfdens}
\Sigma(r_s) = 
\frac{f_\theta }{2 \, f_\Omega \, r_s \, V_c(r_s)} 
\frac{dM}{dt}\; ,
\end{equation}
and drops to zero outside of it.  
Here $f_\theta$ represents the
fraction of the mass that is present along the line of sight compared
to a full spherical shell; at the shell radius $f_\theta = 1$ (otherwise
it would not even be visible), but it can drop off at smaller radii,
as can be seen in the leftmost shell in Figure~\ref{fig.spherecartoon}.
This factor partially compensates for the
solid angle factor $f_\Omega$.  The reason for the constant surface
density at the edge is that as the line of sight moves inward from the shell
edge, it reaches a smaller minimum radius, but also takes a shorter
(increasingly vertical) path through the material at the edge of the
shell; these two effects cancel each other out.  
The relationship just derived allows us to obtain the local mass flow rate
from the surface brightness of the shell, which we make use of in 
\S~4.4.
(\citealp{hernquist88} previously noted that the surface density
approaches a constant, without relating it to dynamical quantities.)

If the shell surface is not tangent anywhere to the line of sight,
as in the rightmost shell in Figure~\ref{fig.spherecartoon},
then the extent of the shell in projection is limited by the angular
coverage rather than the shell radius. The dropoff outside
the shell region need not be sharp, and the boundary of the shell 
on the sky need not be a circular arc.  The surface density in this case 
can be estimated from the average over the sphere.  A single spherical
shell corresponds to a single orbit, so the total mass
contained within the shell equals 
$t_r dM/dt$, where $t_r$ is given by Equation~\ref{eqn.period}.
The total surface area of a full spherical shell on the sky
is $\pi r_s^2$, so the average surface density is 
\begin{equation}
\label{eqn.bodysurfdens}
\Sigma_{avg}(r_s) \approx 
\frac{1}{\pi}
\sqrt{\frac{2}{(-k)}} \, B[1/2,(2-k)/(-2k)] 
\frac{f_\theta}{f_\Omega \, r_s \, V_c(r_s)}  
\frac{dM}{dt}\; ,
\end{equation}
where we have included corrections for the angular coverage as before.
For $k=-0.4$, the constant in front is 0.76, slightly exceeding the
value of 0.5 obtained at a true shell edge (c.f.\
Equation~\ref{eqn.edgesurfdens}), although typically the value of
$f_\theta$ will also be smaller than at an edge.  Thus there is no
strong limb brightening or darkening; the surface density does not
depend very much on whether the line of sight is at a true edge or
not.

\subsubsection{Idealized velocity pattern}

\citet[][henceforth MK]{merrifield98} studied the kinematics of
shells.  They assumed a shell composed of stars on monoenergetic,
low-angular-momentum orbits lying tangent to the line of sight. They
then found that these stars have a characteristic kinematic signature,
which depends only on the shell radius and the total mass contained
within it.  In the space of line-of-sight velocity $v_{los}$ versus
projected radius $R$, the MK result is that stars near the shell radius 
inhabit a region with a triangular boundary:
\begin{equation}
\label{eqn.mk}
| v_{los} | \leq V_c(r_s) \, \frac{r_s - R}{r_s} \; .
\end{equation}
The stars fill this region unevenly, congregating at the boundary.
This boundary is again a caustic feature, which forms because the
velocity component in the line-of-sight direction $\hat{z}$ 
goes to zero at both $z = 0$ and $r=r_s$, so it must reach an extremum 
at some intermediate point along the line of sight.

\begin{figure}
\includegraphics[width=8cm]{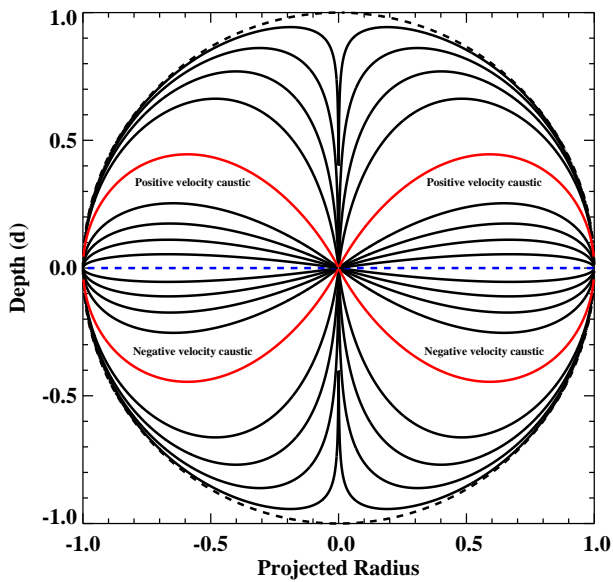}
\caption{
\label{fig.vcartoon}
Location of the velocity caustics, in an idealized model of a spherical
shell of stars.  Here we assume a spherical, power-law potential with
$\Phi \propto -r^{-0.4}$, and a monoenergetic stream.  Projected
radius $R_{proj}$ and line-of-sight distance $d$ are plotted in units of
the shell radius.  We have reflected the pattern around the vertical
axis.  Since the shell geometry is spherically symmetric, rotating
the entire figure around the vertical axis gives the full three-dimensional
pattern.  The shell boundary is shown by the
dashed line.  The contours show the the line-of-sight velocity
relative to the maximum velocity along each line of sight, i.e., at a
fixed $R_{proj}$; contour values are at
($-1$,$-0.8$,$-0.6$...0.6,0.8,1).  The maximum and minimum velocities
along each line of sight occur on the caustics marked with red lines, 
which are labeled appropriately for the outgoing stars.  For the incoming
stars the labels would be reversed.}
\end{figure}

For the discussion below, it is important to understand the relationship
between the position and velocity of the stars.  We illustrate this
in Figure~\ref{fig.vcartoon} for a full spherical shell.  
The region in physical space that corresponds to the velocity caustic
is marked by a red line.  Caustics occur at positive velocities in
two sets of stars: outbound stars lying more distant than the galaxy center,
and inbound stars lying in front of the galaxy center.  There are
negative-velocity caustics made up of the other two combinations.

In the case considered by MK, the velocity distribution is symmetric
around zero, so that the slope is equal and opposite on the positive
and negative velocity sides, and $v_{los} = 0$ at $r_s$.  Also, there
is a degeneracy between the stars on the near and far sides of the
galaxy; outbound stars with $z<0$ have the same sign and distribution
in velocity as inbound stars with $z>0$, and vice versa.  Several effects
may break this degeneracy in a more realistic case, as discussed below.
The method suggested by MK has not yet been applied to measure the 
potential in observed shell
galaxies, due to the low surface brightness of the shells and the
difficulty in estimating the bimodal velocity signature from weak
absorption lines.

\subsection{3-D geometry of the debris in M31}

In the discussion above, we have emphasized the role played by
the angular coverage of the shells.  This is important in our model
since the shell features cover nowhere near the full extent of the
sphere, as can be seen in Figure~\ref{fig.ppos}.  It is worth taking
a closer look at the origin and effects of the spatial structure.

In our simulations, the NE shelf is made up of both material from the
leading stream from the first pericentric passage, which is spread out
into a flattened structure at the second pericentric passage, and material 
stripped from the remnant of the progenitor at its second pericentric passage,
when it completely disrupts.  The disruption is largely a
product of the central density chosen in our initial conditions,
since a more concentrated satellite survived the same passage in
one of our test runs.
While complete disruption is in accord with the lack of an obvious visible
progenitor, it is not {\it required} in order to reproduce the shelves.  
The forward stream is further distorted at the third pericentric passage
into a conical structure, which forms the W shelf.

\begin{figure*}
\includegraphics[width=16cm]{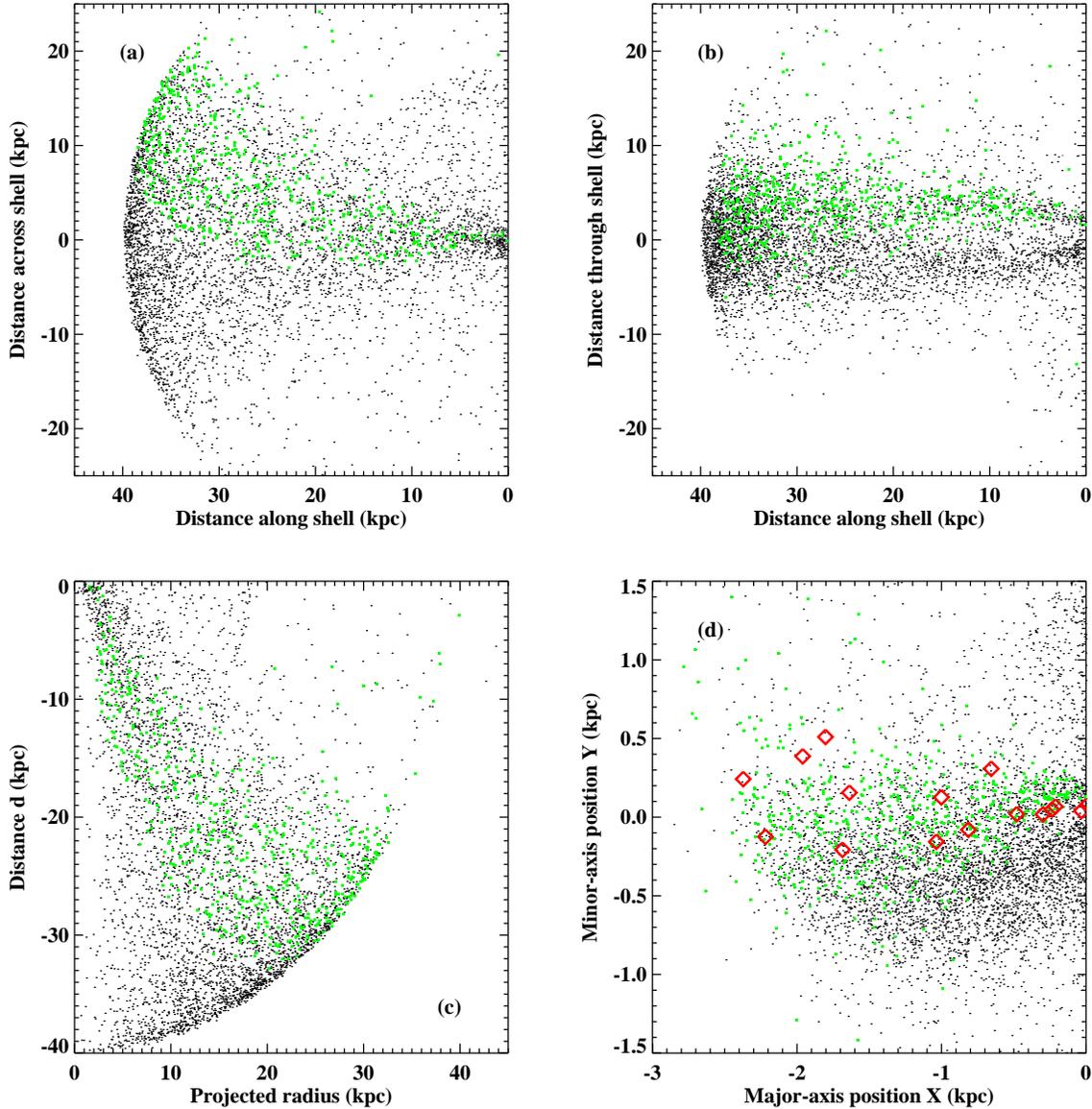}
\caption{
\label{fig.fan}
The NE shelf's fan structure.  
In each panel, the dots indicate $N$-body particles;
those designated as ``Merrett PNe stream'' particles 
in Figure~\protect\ref{fig.xv} are marked in green.
{\it Panel a:} the top view of the structure.  
M31's center is located on the right at $(0,0)$.
The horizontal axis points along the main axis of the structure 
away from M31, while the vertical axis points across the other elongated
direction of the structure.
{\it Panel b:} a side view of the NE shelf structure, at right angles to
the previous panel.  This clearly shows
the outbound (top) and inbound (bottom) layers 
of the double fan structure.  
Again, M31's center is on the right at coordinates $(0,0)$. 
{\it Panel c:} distance $d$ relative to M31, 
versus projected radius on the plane of the sky.
Only those particles in the NE half of M31 ($X<0$) are shown.
Compare the position of the ``stream'' particles here to the
spatial region occupied by the velocity caustic in 
Figure~\protect\ref{fig.vcartoon}.
{\it Panel d:} the sky view of the NE shelf structure, 
plotted in M31 major and minor
axis coordinates ($X$ and $Y$, respectively).
The PNe designated by \citet{merrett06} as
``Stream'' or ``Stream?'' are marked by red diamonds,
and overlay the analogous stream in the simulation.
}
\end{figure*}

The solid angle covered by the simulated NE shelf debris is actually 
even more limited than is apparent in Figure~\ref{fig.ppos}.  When viewed in
three dimensions, it resembles a pair of fans joined at the edges,
corresponding to the outbound and inbound sections; each of these
fan layers is only a few kpc thick, as shown in the second panel
of Figure~\ref{fig.fan}.  
Because this portion of the debris shell does not lie tangent to the
line of sight, its boundary viewed in projection is defined by the
angular coverage and not the shell radius, and thus varies in
projected radius as a function of azimuth.  The latter property
also seems to be shared by the {\em observed} NE shelf
(see Figure~\ref{fig.sky} and Table~\ref{table.edges}), although further
work is warranted to confirm this result.

The southern stream is observed out to near its turnaround radius, so
its length scale is well constrained.  Above we found that the radii
of different shells maintain a fixed ratio with time.  Since the
southern stream is in essence the shell preceding the NE
shelf, this suggests that the shell radii are highly robust for any
model that fits the stream.  Indeed, we found it difficult to
change the shell radii significantly for simulations with acceptable 
stream properties, despite a variety of initial conditions and orbital
phases.

The outbound fan on the E side overlaps the region of the velocity
caustic, as seen in Figure~\ref{fig.fan}.  This leads to a new
interpretation of the Merrett \ea PNe ``stream'': it does not
represent an isolated stream of stars traveling along a single path,
but rather represents an intersection between an approximately two-dimensional
distribution of stars and the mainly two-dimensional region of space
where the velocity caustic occurs.  The various panels of
Figure~\ref{fig.fan} indicate in green the 
``PNe stream'' simulation particles previously marked 
in Figure~\ref{fig.xv}.  
The red symbols in the last panel show the location on the sky
of the ``stream'' PNe found by Merrett et al., which agree well with
the location of our apparent stream.
From the top view of the structure (Figure~\ref{fig.fan}c), however,
it can be seen that the ``PNe stream'' particles do not form an isolated 
density concentration, and actually inhabit a slightly curved
region on the fan structure rather than following a single orbit.
The ``PNe stream'' stars therefore just constitute a portion of a larger
structure, singled out by the kinematics in projection
along the line of sight.

The inbound portion of the fan has its own, inbound ``stream'',
which is offset about $0.5 \degree$ to the SW of the outbound
``stream''.  This unfortunately mainly lies just outside the region
surveyed by Merrett \ea for $X < -1.0 \degree$, and 
for $X > -1.0 \degree$ the velocities mainly lie within the range
expected from disk stars.  However, this returning stream should be 
detectable in a wider PNe survey or a sufficiently large RGB sample.
A sample near $\xi \approx 1.7 \degree$, $\eta \approx 0.7 \degree$
may separate the returning component from the disk spatially, while
a sample near $\xi \approx 1.6 \degree$, $\eta \approx 1.2 \degree$
may separate it kinematically as it should 
have a lower velocity there than that of the disk.  
The existence of a returning stream is a critical test 
of the scenario in this paper,
since it links the outbound NE shelf stars to the W shelf.

The W shelf in our simulation is much less flattened than the NE shelf,
and hence looks more like our idealized notion of a shell in
Figure~\ref{fig.spherecartoon}.  The majority of the stars lie within a
cone $60 \degree$--$90 \degree$ wide, though many of the stars still
lie on a curved concentration on one side of this cone.  The inbound
and outbound stars are intermingled, and some of the inbound stars
have been scattered far outside of the main conical region.  The
stream connecting the E and W stars passes very close ($\ltrsim 1
\kpc$) to M31's center in this simulation, which helps account for the
increased scatter in orbit directions.  The main body of the shelf
lies more distant than M31's center, and is viewed edge-on over much
of its apparent outer boundary, which explains the relatively sharp
and circular appearance in Figure~\ref{fig.ppos}.  

We note that there is yet another radial shell of debris in our
model, consisting of particles even further forward than those in the
W shelf.  This shell is very faint, as it contains only about $6
\times 10^7 \msun$, which is only about a tenth as much mass as the
shell making up the W shelf.  This result is highly model-dependent, 
since this shell represents the extreme forward edge
of the satellite debris.  The shell may not exist at all, or may be
much more massive than found in our simulation.  
The particles in this shell cover the
entire E side of M31 fairly evenly, out to a radius of $18 \kpc$.
Deep kinematic surveys around the SE minor axis would seem to be 
the method most likely to detect this debris.

\subsection{Velocity pattern: expectations and relation to M31's potential}

\begin{figure}
\includegraphics[width=6cm]{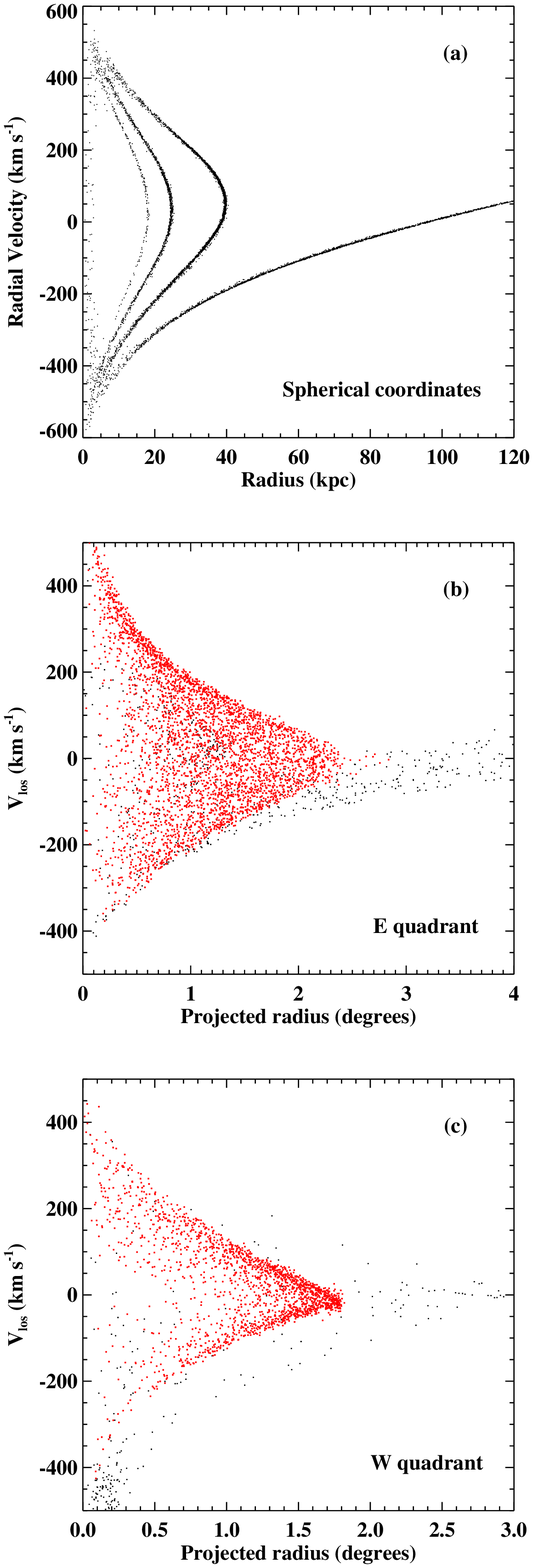}
\caption{
\label{fig.vshelf}
{\it Panel a:} 
M31-centric radial velocity versus M31-centric radius for
the satellite debris particles in the simulation.  The southern, trailing 
stream is made up of the long tail to large radius, while the debris 
further forward spirals clockwise in a series of radial loops.
The NE shelf is made up of the particles in the largest complete loop,
out to 40~kpc,
while the W shelf is made up of the particles in the next loop.  
{\it Panel b:} Line-of-sight velocity
versus projected radius, showing only particles in the 
quadrant $X_{M31} < 0$,
$Y_{M31} < 0$ containing the NE shelf.  Particles genuinely in the
radial shell corresponding to the NE shelf are marked with red
dots.  The black dots show particles belonging
to other radial shells; these are mostly spillover from the southern 
stream, though some debris further forward than the W shelf is also
present at small radii.  Note that in this projection, the 
returning ``stream'' in the NE shelf (the concentration 
at the upper edge of the red particles) 
is more easily seen than in Figure~\protect\ref{fig.xv}.
{\it Panel c:} Same as Panel b, but showing particles 
in the quadrant $X_{M31} > 0$, $Y_{M31} > 0$ containing the W shelf.  
Particles genuinely in the
radial shell corresponding to the W shelf are marked with red
dots. Other particles, mostly spillover from the
NE shelf, are shown as black dots.  
}
\end{figure}

Velocity measurements of the shell debris, as discussed by MK, are
potentially the most powerful constraints on acceptable models of
their makeup.  This 
is partly because velocities can be measured much more accurately than
distances, and partly because the velocities are extremely coherent.
Figure~\ref{fig.vshelf}a shows the location of the particles in the
space of M31-centric radius versus M31-centric radial velocity.  The
particles occupy a cold stream, which spirals in the clockwise
direction as one goes further forward in the stream.  The southern
stream, NE shelf, and W shelf are easily seen as the first three
groups of particles starting from the right.  The fainter
shelf just described on the eastern side is seen as the fourth group.  
The energy decreases along the
stream, reducing the apocentric radii of successive radial loops.

\begin{figure*}
\includegraphics[width=16cm]{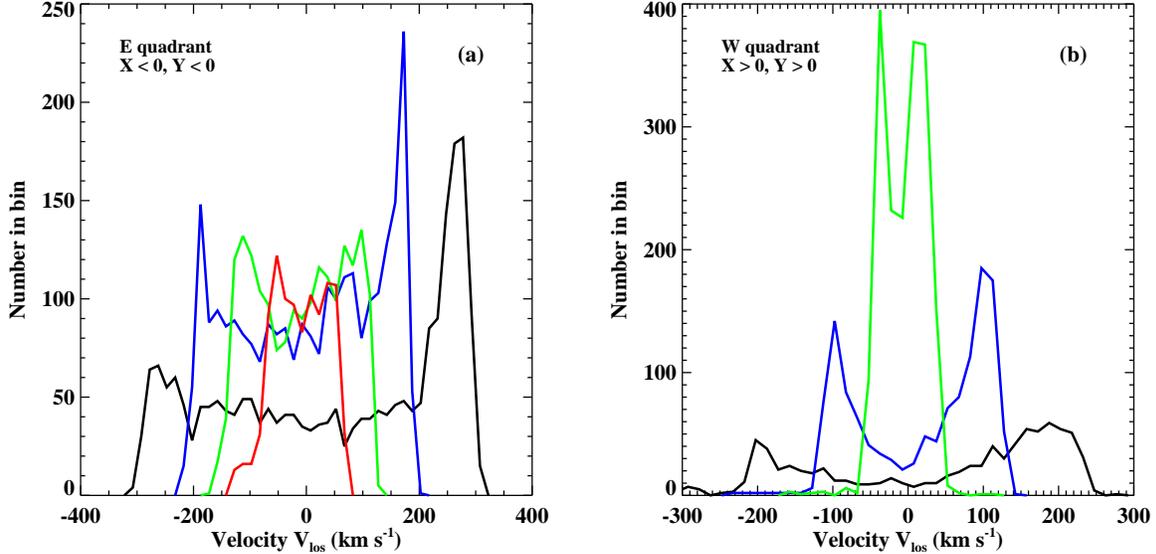}
\caption{
\label{fig.vshelfhist}
Histograms of the line-of-sight velocity of satellite debris
particles at different projected radii.
{\it Left panel:} 
Particles in the quadrant $X_{M31} < 0$, $Y_{M31} < 0$
containing the NE shelf.  Curves show results for
$0.5 \degree< R_{proj} < 0.6 \degree$ (black),
$1.0 \degree< R_{proj} < 1.1 \degree$ (blue),
$1.5 \degree< R_{proj} < 1.6 \degree$ (green),
$2.0 \degree< R_{proj} < 2.1 \degree$ (red).
{\it Right panel:} 
Same, for the quadrant $X_{M31} > 0$, $Y_{M31} > 0$
containing the W shelf. Curves show results for
$0.5 \degree< R_{proj} < 0.6 \degree$ (black),
$1.0 \degree< R_{proj} < 1.1 \degree$ (blue),
$1.5 \degree< R_{proj} < 1.6 \degree$ (green).
}
\end{figure*}

In {\em qualitative} accord with the predictions of MK discussed
above, we find that the velocities lie within a well-defined, roughly
triangular region in the space of $v_{los}$ versus $R_{proj}$, as shown in
Figure~\ref{fig.vshelf} (see also the similar Figure~\ref{fig.wpne}b).  
Over most of the region, the velocity
distribution has two peaks at positive and negative velocity, which
both approach $v \approx 0$ (using velocity relative to M31) as
$R_{proj}$ approaches the shelf boundary.  
These two peaks are shown clearly in the histograms of the
velocity distribution in Figure~\ref{fig.vshelfhist}.  
These two peaks generally have unequal strength, particularly at small
radii in the W shelf where the negative-velocity peak is a fraction of 
its positive-velocity counterpart.
The returning or positive-velocity ``stream'' in the NE shelf shows 
up quite clearly in this representation.  As mentioned above, 
this component has so far eluded detection, probably due mostly
to the areal coverage of the kinematic surveys.

This velocity pattern will be easiest to recognize near the 
edges of the shelf, in the regions far away from M31's disk.  
We discussed evidence that portions of this pattern are visible
in existing observations in \S~3.3, including 
the apparent outbound ``stream'' in the NE shelf.
The corresponding inbound stream is more easily seen in 
Figures~\ref{fig.vshelf} and \ref{fig.vshelfhist}
than in Figure~\ref{fig.xv}, both because the division at $Y = -0.3
\degree$ splits the stream particles between the two panels of the
latter figure, and because the line-of-sight velocity is a function
of $R_{proj}$ rather than $X$, and those two quantities are not
well aligned in the main region occupied by the inbound stream.

We fit the upper and lower envelopes 
in panels (b) and (c) of Figure~\ref{fig.vshelf}
with straight lines, using only the outer portions of each shelf.
Using MK's approximation in Equation~\ref{eqn.mk}, 
we obtain estimated circular velocities of 
$291 \kms$ in the NE shelf at 36 kpc from M31, and
$245 \kms$ in the W shelf at 24 kpc.
The actual circular velocities at these two radii in our potential model
are 214 and 229 kpc, respectively.  Thus, the MK prediction 
is not in precise {\em quantitative} accord with the simulation, but rather
overestimates the enclosed mass by 15--85\%.  There are other disagreements
with the MK description as well: the velocity slopes are not equal and
opposite, and the tip velocity is nonzero.  

Using the simulation particles, we find that these discrepancies originate
from several effects not considered by MK, including the following: 
the nonzero angular momentum in the debris; 
the significant gradient of energy along the stream;
the continuing dynamical interaction with the progenitor, 
mainly in the NE shelf;
perspective effects from the large size and small distance of the M31 system; 
and finally,
the finite range of radius needed to measure the slope.  
All of these effects change the velocities visibly in Figure~\ref{fig.vshelf},
and taken together change the measured slope significantly.
They also break the
above-mentioned symmetry between the caustics from particles on
opposite sides of M31, though the particle distribution is so
heavily weighted to one side in each case
that this effect is barely visible.  
In both shells, the limited spatial coverage of the shell does not
greatly modify the slope of the velocity envelope, although the lack
of stars near the shell edge does deplete the region near the tip and
make the slope measurement more challenging.

Despite these complications, the general point of the MK paper is
still valid: the velocity signature in shells {\em can} serve as a
useful diagnostic of the gravitational potential, over a radial range
15--$35 \kpc$ in this case.  Our gravitational potential is not
necessarily the correct one.  We expect the measurement biases to be
similar for different potentials though, so to first order 
the ratio found here between predicted and measured velocity
gradients can be used to estimate the true enclosed mass at the shelf radii.
Measurements of the H~I rotation curve now probe out to
$\approx 35 \kpc$ as well \citep{carignan06}, but there are significant 
concerns about these results due to the warping and irregular
structure of the disk.  Hence, the shelf velocities will help confirm
the H~I measurements, and will also allow comparison of the potential
in and out of the plane, thereby constraining the ellipticity of the
dark matter distribution.

\subsection{Surface density profile as a diagnostic of the progenitor}

We now investigate what the surface density of the
shelves signifies for the position and mass of the progenitor.
First, we check the approximate relationship between the surface density and
the mass flow rate, using the simulated W shelf as a test case.
Equation~\ref{eqn.edgesurfdens}
is the appropriate form to use for this shelf 
since it is nearly tangent to the line of sight.  
In \S~3.2 we found the simulated surface density in the 
W shelf was $\Sigma_W = 7.2 \times 10^5 \Msun \kpc^{-2}$.
For the other constants in Equation~\ref{eqn.edgesurfdens}
we find $r_s \approx 25 \kpc$, $V_c \approx 230 \kms$, 
and $dM/dt \approx 1.6 \Msun \yr^{-1}$, using the rate of mass in
the shell reaching turnaround as an estimate of the latter quantity.
Converting the particles in this radial shell into a fully
spherical shell by randomizing their orientations 
relative to M31, and computing the surface density in 
the W shelf region as before, we find a surface density of 
$1.5 \times 10^5 \Msun$.  This is about 5 times smaller than 
the actual simulated density, implying
$f_\theta / f_\Omega = 5$ in Equation~\ref{eqn.edgesurfdens},
which is the appropriate form to use for the W shelf 
since it is nearly tangent to the line of sight.  
The expected surface density of the W shelf is then
$\approx 6.3 \times 10^5 \Msun \kpc^{-2}$, 
consistent with the actual surface density in the simulation
within the limits of our accuracy.
Predicting the NE shelf surface density is potentially more complicated, 
since it may contain contributions from both the first and second
pericentric passages of the progenitor satellite, so we will not
make use of it here.

\begin{figure}
\includegraphics[width=8cm]{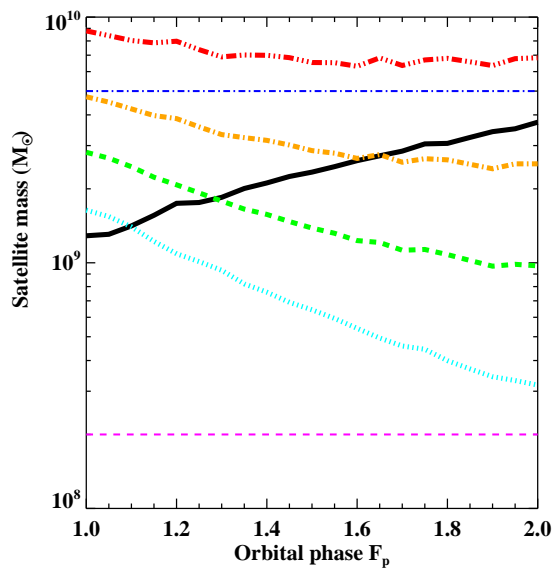}
\caption{
\label{fig.satmass}
Estimates of the progenitor's mass at the time it disrupts
to form the stream,
as a function of the progenitor's current orbital phase.
The orbital phase is defined in units of the progenitor's
radial orbital period, starting from the time of disruption.
The derivations of the various estimates are discussed in the text.  
Rising solid curve: estimate from stellar mass in the stream, accounting
for the mass outside the stream.
Falling curves: estimates from the surface density in the W shelf,
using surface densities of 
2, 4, 8, and $16 \times 10^5 \Msun \kpc^{-2}$
from bottom to top.
Horizontal dashed line: estimate of stellar mass in the stream alone.
Horizontal dot-dashed line: estimate from the metallicity of stream stars
\citep{font06}.  With a better-constrained W shelf surface density, 
the intersection of the rising and falling curves can be used to 
constrain the progenitor mass and current orbital phase.
}
\end{figure}

By viewing successive snapshots of our $N$-body simulation, we can 
easily see that the surface density in the W shell increases with time,
because the mass flow rate in the stream $dM/dt$ is also increasing
as the progenitor approaches closer to this point in the orbit.  This
suggests that we can use the shelf surface brightness to constrain the
progenitor mass and orbital phase.  In \citetalias{fardal06}, we put limits
on the mass and phase using properties of the southern stream alone.
The basic idea was that given a progenitor phase, the progenitor mass
sets both the overall normalization and the phase extent of the stream
mass, and hence the fraction of the mass in the visible extent of the
southern stream.  This argument was limited in its power, though,
because a larger progenitor phase could always be compensated for by a
larger progenitor mass.  With the detection of the W shelf, it seems
that the progenitor phase can be limited from the other direction,
effectively triangulating its position.

We proceed similarly to \S 3.2 in \citetalias{fardal06}.  We fit a sequence of
trajectories assuming different present-day progenitor phases $F_p$,
and otherwise following the method of \S 3.1 above.
For each fit, we calculate 
the time scaling factor $\tau$ at the W shelf 
and at Fields 4 and 8 of \citet{mcconnachie03} within
the southern stream, which span the same radial range as that 
used by \citet{ibata01} in calculating the stream's luminosity.
We also calculate the progenitor's orbital energy $E_0$, 
its progenitor's radial period $t_{r0}$,
its pericentric radius $r_p$,
the W shell radius $r_s$,
and the circular velocity at $r_p$ and $r_s$.

For the distribution of orbital energy in the debris, $\chi(E)$, we
adopt a Gaussian form with the dispersion $\sigma_E$ given in
Equation~\ref{eqn.edispersion}.  This form is a reasonably accurate
approximation to the simulations in \citetalias{fardal06} and this paper, and an
improvement on the top hat distribution used in Fardal et al.  
We estimate the energies at Fields 4 and 8 as
$E_4 = E_0 \tau_4^{-1/3}$ and $E_8 = E_0 \tau_8^{-1/3}$,
using the scalings above with $k=-0.4$.  
We take the total mass of the stream
between fields 4 and 8 as
$M_{\it stream} \approx 2 \times 10^8 \Msun$, as discussed in
\S~\ref{sec.morphology}.
We then estimate the mass of the progenitor satellite $M_s$ 
from that of the stream by solving the following equation:
\begin{equation}
M_s \int_{E_8}^{E_4} \chi(E) \, dE = M_{\it stream} \; .
\end{equation}
This result is shown by the solid curve in Figure~\ref{fig.satmass}.

We then estimate $M_s$ independently from the surface density of the
W shelf using Equation~\ref{eqn.edgesurfdens}.  We relate the mass flow
rate at the edge of the shelf to the steadily increasing energy there
using
\begin{displaymath}
\frac{dM}{dt} = M_s \chi(E) \frac{dE}{dt} \; ,
\end{displaymath}
where the change in energy per unit time $t$ at a fixed orbital phase is
\begin{displaymath}
\frac{dE}{dt} = \frac{dE}{d\tau} \frac{d\tau}{dt} 
  \approx -\frac{E}{3t} \; .
\end{displaymath}
Combining these results with 
Equation~\ref{eqn.edgesurfdens}, we get an equation
that can be solved for the progenitor satellite's mass $M_s$:
\begin{equation}
M_s \frac{dE}{dt}(\tau_{sh}) \chi(E) = 
 2 \frac{f_\Omega}{f_\theta} r_{sh} V_c(r_{sh}) \Sigma(r_{sh}) \; .
\end{equation}
Here we take $f_\theta / f_\Omega = 5$, as 
found using the simulated debris.  
In Figure~\ref{fig.satmass} we 
provide four curves using shell surface densities of $\Sigma_W = 2$, 
4, 8, and $16 \times 10^5 \Msun \kpc^{-2}$, 
which probably cover the allowed range ($8 \times 10^5$ is our 
most likely value, although this includes the smooth halo component
of M31).

The estimate of the progenitor mass and phase is then given by the point
where the W shelf curve, for an observed $\Sigma_W$, crosses the solid
curve from the stream mass.  Both analytic estimates use a series
of approximations and may be imprecise by perhaps a factor of two.
Also, assuming that the stripped fraction or baryonic fraction in the 
progenitor is less than unity will raise both mass estimates together.  
The plot suggests a minimum progenitor mass of 
$M_s \gtrsim 1.5$--$5 \times 10^9 \Msun$, depending on the shelf surface 
density.  These masses are consistent with our simulation and with
the estimate of the progenitor mass from the
metallicity \citep{font06}, shown as the upper horizontal line.
On the other hand, these estimates are far above the integrated 
mass in the stream itself, shown as the lower horizontal line.

The progenitor's orbital phase can already be limited to $F_p \gtrsim 1.1$, 
according to the plot.  The plot also shows that $F_p$
can be determined within a small fraction of a radial period, 
once the surface density in the W shelf is known precisely.

The constraints given here can be improved in the future by using
surface density constraints all the way out to the tip of the southern
stream, and by using an ensemble of $N$-body simulations to avoid the
need for analytic approximations.  Nevertheless, it should be clear
from this discussion that the surface density distribution is an
important clue to the orbital and physical parameters of the
progenitor.

\section{DISCUSSION}

We have performed a simple $N$-body simulation in the gravitational
potential of M31, and used this to show that an accreting satellite
can explain the giant southern stream, the eastern shelf, and
a similar shelf on the western side as a single continuous structure.
Following the entire folded path traced by the stream and shelf
particles, the structure is $\gtrsim 230 \kpc$ long, certainly among
the largest coherent stellar structures known despite its modest total
stellar mass of $\sim 2 \times 10^9 \msun$.  The NE and W shelves in
this model represent successive radial loops of the orbit broadened
into fan shapes by passage through pericenter.
The NE shelf lies closer to us than the disk of M31, and partially
overlaps with it.
This may have relevance to microlensing surveys which aim to detect
compact objects of stellar-type mass \citep{calchi05,dejong04}, 
since the lensing event probability will be boosted relative to disk 
self-lensing by the gap in distance between the disk and NE shelf, 
and by the large relative velocities between the disk and shelf stars.

The counter-rotating ``stream'' in the NE seen in both PNe and RGB
stars in this view represents a continuation of the NE shelf onto the
face of M31, while the apparently narrow velocity width of this stream
is explained by a caustic in velocity space.  The properties of the
incoming satellite are reasonable for a galaxy of its initial stellar
mass, as discussed in \S~3.1.  While we have not explored velocity
distributions outside the shelf regions in detail, our impression is
that the model is satisfactory in terms of where it does not put
debris, as well as where it does.  Our experience with different 
initial conditions (\citetalias[in][]{fardal06}) suggests that this would not
necessarily be true for orbits that tried to associate the southern
stream with other objects such as M32 or the Northern Spur.

We experimented with low-resolution versions of our main run, testing
different orbits and satellite properties.  Our sense acquired during
this exploration is that the volume of parameter space that can match
current observations is large enough not to need extreme fine-tuning.
The time corresponding to ``present day'' within our run is uncertain
by perhaps 100 Myr, and it is not clear whether the progenitor need be
disrupted or not.  As surveys of M31's disk and halo regions continue,
we expect that the requirements on the model will become much more
stringent.

We remind the reader that we have neglected several physical effects in
computing this model that are potentially important.  These include 
dynamical friction, the response of
the modes in the disk, rotation of the accreting
satellite, dark matter in and around its stellar component, 
and the hydrodynamic interaction of gas within the satellite with the
disk and halo of M31.  
In addition, the observations do not yet cover some of
the most interesting areas for this model, and our comparison to the
existing observations is as yet rudimentary.  We therefore regard the
existing simulation as a useful illustration, rather than a definitive 
prediction of the debris structure.  Work on the interaction of 
the satellite with live dynamical components of M31 is 
currently underway.

Kinematic observations in the shelf regions are important for two
reasons.  First, they can help clarify the nature and extent of
the debris related to the giant southern stream.   The many
testable predictions of our model include:
\begin{itemize}
\item The ``PNe stream'' of \citet{merrett06} should be continuously 
connected to the NE shelf, since in our model it is merely a region
where the kinematics in projection gives a roughly constant velocity,
and not a separate structure.
\item In both shelves, the characteristic kinematic pattern of a shell 
should be evident in the $R_{proj}$--$v_{los}$ plane (Figure~\ref{fig.vshelf}).
\item There should be stars with positive velocities returning to the 
center of M31 in the NE shelf, which connect the NE and W shelves.
\item There may be a yet weaker shelf on the E side, from debris further
forward in the stream.
\end{itemize}
Second, the velocity signature of the shells depend on the potential of M31.
Many more spectroscopic fields will be required to allow a precise 
measurement of the potential.  However, unlike the case of the
collimated southern stream, the line-of-sight distances need not be
computed at all to probe the potential of M31, as the geometry itself
selects the stars in the velocity caustic.  This removes one of the main
sources of uncertainty in constraining the mass model.  Comparison of 
the potential derived from the shelves to that obtained from H~I rotation 
curves can help test the deviation of M31's mass distribution from spherical 
symmetry.

If confirmed as shells connected to the southern stream, the NE and W
shelves would to our knowledge represent the first detection
of a multiple-shell system in a late-type spiral galaxy. Shells are very 
common in elliptical galaxies.  137 elliptical galaxies with shells 
were catalogued by \citet{malin83},
while \citet{merrifield98} estimate that 10\% of elliptical
galaxies have shells.
The reasons for this strong morphological bias
have not been completely explained at present. 
Spirals may be destroyed and ellipticals created in the mergers that
produced the shells, although this would only be plausible for 
merging systems much more massive than in the case discussed here
\citep[see][for a discussion of shell formation in major mergers]{hernquist92}.
Alternatively, shells may dissolve faster in the non-spherical potential 
of the disk due to differential precession of the orbits.  
The difficulty of distinguishing shells from spiral features
in late-type galaxies may also contribute to the
lower rate of observing shells \citep{barnes92}.
If all shells around spiral galaxies were as faint as the ones in M31,
they would have escaped detection, since RGB star count maps
like those of \citet{ferguson02} can be made only at close range.
M31's putative shell system would also be the first in
which kinematic observations have been made, allowing us to compare to
and eventually constrain mass models of the host galaxy.  

Theoretical studies of dark halos have established their mean mass
profile to high accuracy in the absence of baryons.  Their response to
the baryonic component in terms of their mass profile, 
ellipticity, and alignment with the disk is more difficult to
simulate, making observational checks important.  However, these
aspects are difficult to study in most galaxies.  The abundance in M31
of both coherent tracers of the potential, such as the tidal debris
discussed in this paper, and statistical tracers such as RGB stars,
satellite galaxies, and globular clusters, suggests that this galaxy
can be a Rosetta Stone for understanding the properties of dark halos
and their interaction with the galaxies they host.

\section*{ACKNOWLEDGMENTS}
We thank Tom Quinn, Joachim Stadel, and James Wadsley for the use
of PKDGRAV, and Josh Barnes for the use of ZENO.
Helen Merrett and the PN.S team graciously supplied 
their table of PNe data prior to publication.  
We thank Jason Kalirai, Scott Chapman, Mike Irwin, Tom Brown,
Roger Davies, and Jon Geehan for helpful conversations.
MF is supported by NSF grant AST-0205969
and NASA ATP grants NAGS-13308 and NNG04GK68G.
PG acknowledges support from NSF grants AST-0307966 and
AST-0507483 and NASA/STScI grants GO-10265.02 and GO-10134.02.
Research support for AB comes from the Natural Sciences and
Engineering Research Council (Canada) through the Discovery and the
Collaborative Research Opportunities grants.  AB would also like to
acknowledge support from the Leverhulme Trust (UK) in the form of the
Leverhulme Visiting Professorship.  
AWM thanks Sara Ellison and Julio Navarro for financial support.

\label{lastpage}
\end{document}